\documentclass[hyper,12pt,letterpaper]{JHEP3}

\newcommand{\eq}[1]{(\ref{#1})}

\newcommand{\EE}{{\cal E}}
\newcommand{\NN}{{\cal N}}
\newcommand{\MM}{{\cal M}}

\newcommand{\wt}{\widetilde}

\title{Black Hole Microstates and Attractor Without Supersymmetry}

\author{Atish Dabholkar$^1$, Ashoke Sen$^2$ and
Sandip P. Trivedi$^1$ \\
~$^1$Tata Institute of Fundamental Research, \\
Homi Bhabha Road, Mumbai 400005, India \\
~$^2$Harish-Chandra Research Institute,  \\
Chhatnag Road., Jhunsi,
Allahabad 211019, India
\\
}

\abstract{Due to the attractor mechanism, the entropy of an extremal
black hole does not vary continuously as we vary the asymptotic
values of various moduli fields. Using this fact we argue that the
entropy of an extremal black hole in string theory, calculated for a
range of values of the asymptotic moduli for which the microscopic
theory is strongly coupled, should match the statistical entropy of
the same system calculated for a range of values of the asymptotic
moduli for which the microscopic theory is weakly coupled. This
argument does not rely on supersymmetry and applies equally well to
nonsupersymmetric extremal black holes. We discuss several examples
which support this argument and also several caveats which could
invalidate this argument.}


\preprint{TIFR/TH/06-31\\ HRI-P-06-11-002 \\ hep-th/0611143}

\newcommand{\p}{\partial}
\newcommand{\LL}{{\cal L}}

\newcommand{\half}{\frac{1}{2}}

\newcommand{\apm}{\alpha'}

\def\e{\epsilon}

\def\l{{\lambda}}

\def\CN{{\cal N}}

\def\half{{\frac12}}

\def\IC{\relax\hbox{$\inbar\kern-.3em{\rm C}$}}
\def\bZ{{\bf Z}}

\def\bR{{\bf R}}

\def\tQ{{\wt Q}}

\def\IC{{\bf C}}

\def\CN{{\cal N}}

\def\bea{\begin{eqnarray}}
\def\eea{\end{eqnarray}}
\def\be{\begin{equation}}
\def\ee{\end{equation}}
\def\ba{\begin{align}}
\def\ea{\end{align}}
\def\bse{\begin{subequations}}
\def\ese{\end{subequations}}
\def\1F1{{}_1\!F_1}
\def\2F0{{}_2\!F_0}

\def\bP{$\bar{\rm P}$\,}

\begin{document}
\setcounter{tocdepth}{2}

\section{Introduction}

One of the important successes of string theory is that one can
obtain a statistical understanding of  the thermodynamic
Bekenstein-Hawking entropy of certain supersymmetric black holes in
terms of microscopic counting\cite{Strominger:1996sh}. The main
theoretical  tool in much of this work is the BPS property of these
supersymmetric black holes. A BPS state in theories with ${\cal N}
=2$ or more supersymmetry belongs to a short representation of the
supersymmetry algebra. As a result, under suitable conditions, the
number of BPS states cannot jump discontinuously under smooth
variations of  the coupling constant and other moduli. The spectrum
of BPS states with a given assignment of charges can then be
reliably computed at weak coupling and then analytically continued
to the strong coupling regime where the same state is described by a
supersymmetric black hole. This allows us to compare the statistical
entropy with the Bekenstein-Hawking entropy even though they are
calculated in different regions in the coupling constant space. In
addition the BPS property of these states leads  to considerable
computational simplification. Exact solutions describing the
corresponding black hole  in supergravity can be found by solving
first-order Killing spinor equations instead of second-order
equations of motion.

A further significant simplification  results from the `attractor
mechanism' noted first in the context of supergravity
\cite{Ferrara:1995ih, Strominger:1996kf, Ferrara:1996dd} and
generalized to theories with higher derivative terms in
\cite{LopesCardoso:1998wt, LopesCardoso:1999cv, LopesCardoso:1999ur,
LopesCardoso:2000qm}. The moduli fields in a black hole background
vary radially and get attracted to certain specific values at the
horizon which depend only on the quantized charges of the black hole
under consideration. As a result, the macroscopic entropy is
determined purely in terms of  charges and is independent of the
asymptotic values of the moduli. This is consistent with the fact
that the microscopic entropy is also independent of the
asymptotic moduli due
to the BPS property of the state it counts. The attractor values of
the moduli are determined by solving a set of `attractor equations'
which are purely {\it algebraic}. Thus, with the attractor
mechanism, the problem of finding the entropy of supersymmetric
black holes is simplified enormously and reduced to solving
algebraic equations instead of non-linear second or higher order
differential equations.

Using the generalized attractor mechanism, and using the proposal
for mixed statistical ensemble proposed in \cite{Ooguri:2004zv},
it has recently become possible to carry out a far more detailed
comparison between microscopic and macroscopic entropy. For a
number of examples of both small and large black holes the two
entropies agree  to all orders in a perturbation theory in inverse
charges going well beyond the thermodynamic
Bekenstein-Hawking result
\cite{Dabholkar:2004yr, Dabholkar:2004dq, Sen:2004dp, Sen:2005pu,
Dabholkar:2005by,Dabholkar:2005dt}.

Much of this success is crucially tied to supersymmetry and it is
interesting to ask if some generalization to non-supersymmetric
black holes is possible. Indeed, there are already a number of
indications that the attractor mechanism as well as the agreement
between thermodynamic and statistical entropy could work even
without supersymmetry for extremal black holes.

\begin{itemize}
\item  For many extremal but non-supersymmetric black holes within
string theory, both in four and five dimensions, the macroscopic
entropy agrees with the microscopic degeneracy of states computed
at weak coupling \cite{Kaplan:1996ev, Horowitz:1996ac,
Dabholkar:1997rk,Emparan:2006it}.
Such an agreement is {\it a priori}
quite mysterious, because
these black holes are not `nearly supersymmetric' in any sense and
break supersymmetry completely. Since they belong to a long
representation of the supersymmetry algebra, one cannot invoke the
argument given above for the analytical continuation of their
spectrum from weak coupling to strong coupling
in an obvious way.

\item The attractor mechanism is a consequence not so  much of
supersymmetry but rather of the near horizon extremal geometry which
is $\bf AdS_2 \times S^n$  in $n+2$ dimensions. For a general class
of two derivative actions describing gravity coupled to scalar
fields and abelian gauge fields, extremal black holes are known to
exhibit attractor phenomenon under certain conditions even without
supersymmetry
\cite{Ferrara:1997tw,Sen:2005wa,Goldstein:2005hq,Kallosh:2005ax,
Tripathy:2005qp,{Kaura:2006mv}, Giryavets:2005nf,Goldstein:2005rr,
Kallosh:2006bt,Kallosh:2006bx,{Khuri:1995xq}, {Ortin:1996bz}}. For
the non-supersymmetric  black holes mentioned above, where exact
supergravity solutions are known, several moduli do get attracted to
fixed values at the horizon irrespective of their values at
asymptotic infinity.

\item For a completely general class of gravity actions including
arbitrary higher derivative interactions, assuming an extremal
near horizon geometry but without assuming supersymmetry, the
attractor values of moduli can be obtained by extremizing an
`entropy function'. The value of the function at the extremum gives
the full Bekenstein-Hawking  entropy of these black holes
\cite{Sen:2005wa,Sen:2005iz} after
inclusion of higher derivative corrections
to the entropy formula following
Wald's
procedure\cite{Wald:1993nt,Iyer:1994ys,Jacobson:1993vj,Jacobson:1994qe}.
\end{itemize}

Motivated by these results, we investigate the question of the
microscopic interpretation of the entropy of non-supersymmetric but
extremal black holes within string theory. In  $\S{\ref{attractor}}$
we propose an argument as to why the  microscopic and macroscopic
entropy of an extremal black hole should agree despite the fact that
they are calculated in different regimes in the coupling constant
space. Our argument does not rely on supersymmetry but relies rather
on the attractor phenomenon. The basic underlying idea is the
following.

While in absence of supersymmetry we lack an argument that allows us
to continue the expression for the statistical entropy from the weak
coupling to the strong coupling regime, the attractor mechanism, --
which tells us that the entropy of an extremal black hole does not
change as we vary the asymptotic values of the moduli fields, --
allows us to continue the expression for the black hole entropy from
the strong coupling to the weak coupling regime where it can be
compared with the statistical entropy. Based on this argument we
present a conjecture that for all extremal black holes, the
macroscopic entropy will agree with the weak-coupling
microscopic entropy as long as certain conditions are satisfied. In
particular the geometry must approach $\bf AdS_2\times S^n$ form
near the horizon which can be modified but not destabilized by
higher derivative corrections, an interpolating solution must exist
that connects the weakly coupled region at asymptotic infinity to
the attractor geometry near the horizon and the near horizon field
configuration should not jump discontinuously under a continuous
variation of the asymptotic moduli from strong to weak coupling
regime. Besides providing new information on extremal
non-supersymmetric black holes, our argument also provides a new
explanation of why the statistical and black hole entropy agree for
extremal BPS black holes.

One subtlety that arises in the comparison between statistical
entropy and the black hole entropy involves the precise definition
of the statistical entropy of extremal black hole. Since the mass of
a non-BPS state can change continuously as a function of the
coupling, the degeneracy of strictly lowest energy states in a given
charge sector may change as we vary the coupling. A more appropriate
definition would be the logarithm of the total number of states
within a given range of the lowest energy eigenvalue, or
equivalently the statistical entropy calculated for a small but
non-zero temperature. On the other hand there is also a potential
problem in defining the entropy of a strictly extremal non-BPS black
hole due to the fact that some of the flat directions of the leading
entropy function could be lifted due to higher derivative terms in
the action, and the full entropy function may not have a non-trivial
extremum. In this case we shall have a runaway behavior of the
moduli fields as we approach the horizon of an extremal black hole,
and we need to control this by introducing a small amount of
non-extremality.
Both these issues as well as their relationship are
discussed in \S\ref{discussion}.

Once we introduce a small amount of nonextremality, the entropy is
no longer strictly independent of the asymptotic moduli.
Thus the validity
of our argument will depend on the extent to which the entropy
begins to depend on the `flat directions'.
We need to analyze
the dynamics on a case by case basis to settle this issue.
Often it is possible, based on other arguments, to
determine the order at which the equations of motion associated with
the flat directions begin receiving a non-trivial contribution. The
entropy function formalism then tells us that the dependence of the
entropy on the flat directions also begins at that order, -- the
point being that the function whose extremization gives the
equations of motion is the same function whose value at the extremum
gives the entropy. If the order at which this dependence begins
remains subleading as we vary the moduli from the
strong coupling to the weak coupling regime,
then our argument about the equality of
microscopic and macroscopic entropy remains valid. An example
of such a situation can be found in \S\ref{srunaway}.

In $\S{\ref{blackcount}}$ we review various known examples of
non-supersymmetric extremal black hole
solutions\cite{Dabholkar:1997rk,Kaplan:1996ev,Tripathy:2005qp} where
the microscopic entropy is known to agree with the macroscopic
entropy despite lack of supersymmetry. In
$\S{\ref{interpolating}}$ we show in
detail how our argument works for a
specific example of five dimensional black hole described in
$\S{\ref{blackcount}}$. In \S\ref{srunaway} we explore,
with the help of some examples,
what happens if some of the marginal directions of the
entropy function get lifted after inclusion of higher derivative
terms and the resulting entropy function does not have an extremum.
In this case for a strictly extremal black hole the geometry and
other background fields keep evolving as we go down the infinite
throat of the would be $\bf AdS_2$ and we never reach the near horizon
$\bf AdS_2\times S^n$ geometry. However we show that by introducing a
small amount of non extremality we can tame this runaway behavior
and get a black hole solution sufficiently close to the original
extremal black hole solution in the absence of higher derivative
terms. Thus the entropy function method can be used to calculate the
entropy of such black holes. Furthermore our argument showing the
independence of the entropy of the asymptotic moduli will hold for
the entropy of such black holes to a good approximation.

Most of the analysis in this paper
is based on the assumption that the near horizon
geometry of the black hole has an $\bf AdS_2$ factor. In many examples
in string theory black holes one finds that this $\bf AdS_2$ factor combines
with an internal compact circle to produce a locally
$\bf AdS_3$ space. In such cases the additional symmetries of
$\bf AdS_3$ allows us to derive results which are much more
powerful than the ones based on the assumption of only the
$\bf AdS_2$
geometry\cite{Kraus:2005zm,Kraus:2005vz,Kraus:2006wn}. In
section \S\ref{scomp} we review the results obtained
using the assumption of $\bf AdS_3$ near horizon geometry, and also
discuss the relative strength and weakness of
this approach compared to the $\bf AdS_2$ based
approach. In section \ref{srotate} we generalize our analysis to
include the case of extremal rotating black holes.

\section{Microstate Counting and the Non-supersymmetric Attractor}
\label{attractor}

In this section we shall argue that subject to certain conditions
being satisfied, the microscopic entropy of an extremal black hole
must match the macroscopic entropy even in the absence of
supersymmetry. The issue at hand is the following. Let us take all
the non-zero charge quanta to be large and (say) of the same order
$N$ and let  $\lambda$ be the closed string coupling constant. The
microscopic entropy of this system can be  calculated in the range
of $\lambda$ where we can describe the dynamics of the system in
terms of a set of weakly interacting degrees of freedom. Typically
this requires a combination involving positive powers of $\lambda$
and $N$ to be small, {\it e.g.} for a D-brane system this requires
the 't Hooft coupling $\lambda N$ to be small. We shall call this
the weak coupling region of the moduli space. But in this region
gravity is weak and the horizon of a would-be black hole carrying a
fixed set of charges form at such a small radius that the classical
supergravity description breaks down at the horizon. Hence there is
no conventional black hole solution describing the system. If we now
keep the charge quanta fixed but increase the coupling constant,
then the horizon radius of the would be black hole would grow and
eventually we get a regular black hole solution. In this region we
can reliably  calculate the black hole entropy, but the microscopic
degrees of freedom become strongly interacting and hence we cannot
reliably compute the microscopic entropy. We shall call this the
`strong' coupling region.\footnote{For large $N$ this can be done by
keeping $\lambda$ small so that the asymptotic theory is still
weakly coupled. Thus by `strong' coupling region we shall mean that
the microscopic degrees of freedom of the black hole are strongly
coupled but the asymptotic theory is weakly coupled. } The question
is: How can we compare the two entropies calculated in two different
regions in the coupling constant space?

For supersymmetric states the BPS condition allows us to
analytically continue the expression for the statistical entropy
computed for weak coupling into the regime of `strong'  coupling. This
analytic continuation is justified by the classic argument  of
Witten and Olive\cite{Witten:1978mh} that relies on the fact that a
BPS state belongs to a short representation of the supersymmetry
algebra and hence the number of BPS states cannot jump
discontinuously as we continuously vary the parameters of the
theory. Thus if one had a similar argument for the
non-renormalization of the degeneracy of states for the non-BPS
states, then we could continue the answer for the statistical
entropy from weak coupling region to `strong' coupling region, and
compare this with the black hole entropy. Unfortunately such a
non-renormalization theorem is not available for the statistical
entropy of non-BPS states.

This is where the attractor mechanism comes to our rescue. This
allows us to run the argument backwards, -- namely we calculate the
black hole entropy in the `strong' coupling region, and then continue
the result to the weak coupling region using the fact that the black
hole entropy is independent of the asymptotic value  of the string
coupling constant $\lambda$. In the weak coupling region we can
compare the result with the statistical entropy.

Let us elaborate on this point in some detail. We can view the black
hole geometry as an interpolating geometry from the asymptotic
infinity to the horizon. At large coupling the curvatures are small
everywhere in the geometry. Thus we can calculate the entropy of the
black hole as a systematic expansion in inverse powers of $N$ using
Wald's formula or equivalently the entropy function defined in
\cite{Sen:2005wa,Sen:2005iz}. For small   coupling, as we move
radially inwards, the spacetime will typically develop regions of
high curvatures. In these regions, it would be necessary to go
beyond the supergravity approximation and include the higher
derivative corrections to the low energy effective action. We can
formally include all higher derivative corrections keeping all terms
in the effective action. Then \emph{assuming} that  the fully
corrected spacetime geometry exits into an ${\bf AdS_2\times S^n}$
geometry (possibly with large curvature or large coupling constant)
as we move radially inwards, one can formally compute the full Wald
entropy using the entropy function that incorporates the effects of
the higher derivative terms. The parameters labeling the near
horizon field configuration are obtained by extremizing the entropy
function with respect to these parameters, and the entropy is given
by the value of the entropy function at this extremum. If the
entropy function has a unique extremum, then of course the near
horizon field configuration and the entropy are uniquely determined
by the entropy function and cannot depend on the asymptotic moduli.
If the entropy function has one or more flat directions then not all
the moduli at the horizon are determined in terms of the charges and
could depend on the asymptotic values of the moduli fields. However
the entropy, being the value of the entropy function at the
extremum, will not depend on the asymptotic
moduli\cite{Sen:2005wa,Sen:2005iz}. Thus the final entropy will have
the same value for `strong' and weak coupling and the entropy will
continue to have the same perturbative expansion in inverse powers
of $N$ where $N$ stands for some typical charge of the black hole.

We can present the argument in another way that does not directly refer
to having a near horizon $\bf AdS_2\times S^n$
geometry at weak coupling. Let
us denote by $f(\lambda)$ the black hole entropy as a function of
$\lambda$.  Now the analysis based on the entropy function tells
us that for large $\lambda$ it is strictly independent of $\lambda$
provided the contribution of the higher derivative terms do not
destabilize the $\bf AdS_2\times S^{n}$ near horizon geometry.
If we now
assume further that $f(\lambda)$  is an analytic function
of $\lambda$, then
it must be strictly independent of $\lambda$ in the full complex
$\lambda$ plane, or a region in the complex $\lambda$ plane containing
the `strong' coupling region  in
which $f(\lambda)$ is analytic. If this region includes the weak coupling
region then $f(\lambda)$ in the weak coupling region will have the
same value as in the `strong' coupling region.

At this point special mention must be given to small black holes --
black holes which describe elementary string excitations. In this
case there is no regular horizon in the supergravity approximation;
the closest analog to the attractor geometry is a scaling region
where the solution becomes independent of all asymptotic
parameters\cite{Sen:1995in,Peet:1995pe}. Analyzing the behavior of
the solution in this scaling region and knowing certain general
structure of the string effective action one can show that up to an
overall normalization factor that is not determined by the scaling
argument, the entropy of the small black hole agrees with the
statistical entropy computed from the elementary string spectrum
\cite{Sen:1995in,Peet:1995pe,Sen:2005kj}. Further analysis based on
certain non-renormalization theorem then shows that the overall
normalization constant also
agrees\cite{Dabholkar:2004yr,Kraus:2005vz}. Given that the
supergravity solution does not have a regular horizon one might
wonder about the relevance of our argument in the context of small
black holes. To this end we note that in order that the solution
enters the scaling region we need to adjust the asymptotic coupling
so that we are in the `strong' coupling region in the sense
described above. Otherwise before we enter the scaling region the
curvature and other field strengths become strong. We then need to
invoke the independence of the entropy on the asymptotic parameters
to argue that the black hole entropy remains the same as we go to
the weak coupling region.

These arguments are predicated on several important assumptions which
we list below:
\begin{enumerate}

\item We assume
 that after
including the higher derivative corrections, the near horizon
geometry still is of the form $\bf AdS_2 \times S^{n}$ so that
we can apply the formalism of \cite{Sen:2005wa,Sen:2005iz}.
Note that it does not require a
detailed knowledge of the interpolating geometry, and not even the
complete details of the near horizon field configuration but only that it
exits into a near  horizon attractor geometry of the form $\bf
AdS_2 \times S^{n}$. Experience with small and large black holes
indicates that this assumption is likely to be satisfied at least
in a large number of cases.  In
fact in the case of small black holes
 the higher derivative corrections actually create the
  $\bf AdS_2 \times S^{n}$ near
horizon geometry even though in the
supergravity approximation the geometry is
singular\cite{Dabholkar:2004yr, Dabholkar:2004dq}. In general it is
quite difficult to analyze the details of the full geometry reliably
once the curvatures are large unless there is some help from
supersymmetry. However, the entropy in many examples appears to be
more robust than other inessential details of the geometry.

The arguments based on analyticity bypasses the need of having
$\bf AdS_2\times S^n$ near horizon geometry in the weak coupling
region, but it requires existence of $\bf AdS_2\times S^{n}$ geometry
in the `strong' coupling region even after inclusion of all the higher
derivative corrections. As we shall discuss, this may not always
be true if some of the flat directions of the entropy function are
lifted after inclusion of higher derivative terms and the resulting
entropy function has no extremum.

\item A key ingredient in our argument is the fact that for an
$\bf AdS_2\times S^n$ near horizon geometry the entropy does not change
as we continuously vary the asymptotic coupling constant. This in
turn follows from the fact  that the black hole entropy is obtained
by extremizing an entropy function with respect to the parameters
labeling the near horizon geometry.
For a
local action there is a well defined algorithm
for constructing the entropy function from the local
Lagrangian density\cite{Sen:2005wa,Sen:2005iz}.
But typically fully quantum corrected effective
action
has non-local terms and it is not {\it a priori}
guaranteed that the
notion of
entropy function will continue to hold in the presence of
such terms.
 In our argument we have implicitly assumed
that the entropy function formalism
 continues to hold for full quantum corrected effective
action which could in principle contain non-local terms as well.
This assumption is essential in cases where quantum corrections to
the effective action are important in the near horizon geometry of
the black hole.\footnote{Even if the string coupling is small at the
horizon, some other parameters, {\it e.g.} inverse sizes of the
compactification manifold, may become large, forcing us to use a
dual description. In this dual description the string coupling may
not be small.}

\item Even if both the above assumptions are correct,
a discontinuous change in the entropy may arise as we vary $\lambda$
if $\lambda$ crosses over to a different basin of attraction.
Typically this will move the near horizon geometry to a different
extremum of the entropy function and will change the value of the
entropy. Clearly our argument will break down if this happens.

\item Typically in the supergravity approximation the entropy function
has several flat directions both for BPS and non-BPS extremal black
holes. Once higher derivative corrections are taken into account
some of these flat directions may be lifted. Generically for
supersymmetric black holes there are non-renormalization theorems
which  prevent this, but there is no such result for
non-supersymmetric black holes. If the resulting entropy function
has an extremum where the curvatures and other field strengths are
small we can still calculate the entropy function in a systematic
expansion in inverse powers of $N$, and higher derivative terms will
give rise to small corrections to the leading entropy. However there
could be potential problem if the resulting entropy function has no
extremum. In this case if we follow the radial evolution of various
fields, there will be a runaway behavior as we approach the horizon
and we shall not get an $\bf AdS_2\times S^n$ near horizon geometry.
Even if there is an extremum but at the extremum the near horizon
geometry has large curvature where the higher derivative corrections
are important, then there can be large correction to the leading order
result for the entropy.
As
a result even at `strong' coupling,  when the curvature is small
everywhere in the supergravity approximation, higher derivative
corrections will modify the solution in a non-trivial way that would
seem to completely invalidate the leading order result.

One way to avoid this problem is to consider slightly non-extremal
black holes instead of exactly extremal black holes.
In this case the near horizon geometry is no longer
$\bf AdS_2\times S^n$, but for sufficiently large charges and small
extremality parameter there will be a long throat region
 where the geometry will be approximately $\bf AdS_2\times
 S^n$. We can then calculate the approximate value of the
 entropy by evaluating the entropy function in this region.
For our argument to be valid, we need to assume that the entropy of
such a black hole remains approximately independent of the asymptotic
values of the moduli fields all the way from the strong coupling to the
weak coupling region.
 This issue together with its microscopic counterpart will be discussed
 in more detail in \S\ref{discussion}, and will be illustrated with
 example in \S\ref{srunaway}.

\item Another important assumption that has gone into our
argument is the identification of the extremal black hole with the
lowest mass state for a given set of charges. As explained above, an
extremal black hole is defined by the requirement that its near
horizon geometry is $\bf AdS_2\times S^{n}$. The entropy
function formalism allows us to compute the entropy of these black
holes for a given set of charges but does not give us any
information about its mass. On the other hand when we compute the
degeneracy of states by identifying the black hole with a
configuration of branes in string theory, we typically calculate the
degeneracy of states with the lowest mass consistent with a given
set of charges. In our argument we have implicitly assumed that
these two requirements are identical, \i.e. an extremal black hole
always describes the lowest mass state with a given set of charges.
This is of course true when the space-time curvature is small
everywhere outside the black hole horizon, but may break down when
there are regions of strong curvature in the black hole solution.

\item A related issue is that
of a precise definition of statistical entropy of an extremal black
hole. In the case of supersymmetric black holes
there is a clear distinction between BPS states and nearly
BPS states since they belong to different representations of the
supersymmetry algebra. Thus we can define the statistical entropy of
BPS states by counting the number of BPS supermultiplets. But
in absence of supersymmetry there
is no such clear distinction between the lowest mass states and
other states and it would seem more natural to
define the statistical entropy as the logarithm of the total number
of states with mass within a small range of that of the lowest mass
state. We shall discuss this point in more detail in
\S\ref{discussion}. For the time being we note that this
fits in well with the requirement of introducing a small amount of
non-extremality in the black hole description due to lifting of the
flat directions since
the latter corresponds
to introducing a small temperature
or equivalently defining the
entropy by counting the total number of states within a small
energy range around the lowest energy state.

\item In our analysis we have assumed that the black hole under
consideration is stable. For BPS black holes this follows as a
consequence of supersymmetry, but this need not be true for non-BPS
black holes. Nevertheless we expect that as long as the black hole
does not have any classical instability, it should at least be long
lived (if not stable) since there is no Hawking radiation from
extremal black holes and we should be able to define the entropy of
such black holes. On the microscopic side the corresponding
microstates should also be long lived since they are the lowest mass
single particle states for a given charge, and hence the phase space
available to them for decaying into lower mass particles should be
small. Hence it should be possible to define the entropy on both
sides and carry out the comparison of the black hole entropy with
the statistical entropy. Notwithstanding these general arguments,
stability of extremal non-supersymmetric black holes clearly is an
issue that should be examined in detail on a case by case basis. Our
arguments will apply only to the cases where the black hole is stable
or long lived.

 \end{enumerate} Subject to these caveats, our
arguments suggest the following conjecture.

\noindent \textbf{Conjecture:}\emph{Thermodynamic entropy of
extremal black holes  in string theory matches with the statistical
entropy determined by counting of underlying microstates at weak
coupling. }

\noindent
This conjecture says that  the attractor mechanism in
effect provides a non-renormalization theorem for the degeneracy of
states which carry the lowest mass for given charge.

In \S\ref{blackcount} and \S\ref{interpolating} we will elaborate on
this argument through various examples.

\section{Defining the  Entropy of Non-BPS Extremal Black Holes}
\label{discussion}

If our conjecture is correct in full generality, then the reasons
for the agreement between macroscopic and microscopic entropy appear
to go well beyond the usual arguments from BPS stability. In this
section we address the question of precise definition of the
microscopic and macroscopic entropy that goes
into the aforementioned
correspondence.

First let us consider the case of BPS black holes.
Our conjecture  implies that the macroscopic entropy of an extremal
black hole should agree with the weak-coupling statistical entropy.
By definition, statistical entropy is always the logarithm of the
absolute number of microstates carrying a given set of macroscopic
charges. However, often in comparing the statistical and black hole
entropy for BPS states one uses an index rather than the absolute
number to compute the statistical entropy. The rationale behind this
is the underlying assumption that in general at `strong' coupling
whatever states could combine with other states to become non-BPS
will do so, and only the index worth of states will remain in the
spectrum of BPS states. Thus at `strong' coupling the absolute number
of microstates is equal to the index. There are some notable
exceptions to this rule; the simplest examples being the ones
discussed by Vafa in \cite{Vafa:1997gr} for supersymmetric black
holes. In many cases discussed there the absolute number of black
hole microstates with three charges scales as $N^{3/2}$ in agreement
with the entropy whereas the index scales as $N$.
Except for this ambiguity, the statistical entropy of a BPS black
hole is well defined, since a BPS state can be clearly distinguished
from a non-BPS state by its supersymmetry transformation
property.

The definition of macroscopic entropy of a BPS black hole is also
reasonably clean. The attractor phenomenon tells us that the black
hole entropy does not vary continuously as we vary the asymptotic
moduli. In particular if the near horizon values of some moduli are
not determined by the attractor equations, then the entropy is
independent of these moduli. We also expect that in many (if not
all) cases supersymmetry will prevent lifting of these flat
directions by higher derivative terms and associated runaway
behavior, especially if the near horizon geometry has enhanced
supersymmetry as in \cite{LopesCardoso:1998wt,LopesCardoso:1999cv,
LopesCardoso:1999xn,LopesCardoso:2000qm}. Hence the entropy of such
black holes remains well-defined.

For non-BPS black holes the situation is much more murky. First of
all, on the microscopic side there is no analog of an index, and
there is no clear distinction between the lowest energy state and
nearby states with slightly higher energy. Even if the lowest energy
state is degenerate at zero coupling, once a small coupling is
switched on the degeneracy may be lifted unless it is protected by
some symmetry. This suggests that a more appropriate quantity will
be the total number of states which are within a small but fixed
mass range $\epsilon$ or equivalently the entropy calculated at a
small but non-zero temperature. For small enough coupling when the
correction to the mass of a state is smaller than the parameter
$\epsilon$, the statistical entropy calculated at zero coupling can
be expected to be equal to that calculated at weak coupling. The
entropy defined this way however acquires a subleading piece that
depends on the precise nature of the energy cut-off as well as on
the various moduli characterizing the vacuum.

Apparently independent of these considerations, the possible runaway
behavior at the horizon, associated with lifting of the flat
directions of the entropy function by the higher derivative
corrections, may require us to introduce a slight amount of
non-extremality on the black hole side. To see how it works, let us
denote by $\epsilon$ the non-extremality parameter. The effect of
the non-extremality parameter is to truncate the infinite throat of
$\bf AdS_2$ into a finite size, and as a result the near horizon
geometry is no longer $\bf AdS_2\times S^n$. Since the original
runaway behavior came from radial evolution along the infinite
throat of the $\bf AdS_2$ geometry, we expect that for any finite
$\epsilon$ various fields will approach finite values at the horizon
instead of showing runaway behavior. However for sufficiently large
charges and small $\epsilon$ there will be a region in the black
hole space-time where the geometry is approximately $\bf AdS_2\times
S^n$, and we can apply the entropy function formalism to calculate
the entropy in this region. (This will be demonstrated in
\S\ref{srunaway} with the help of some examples.) Although this does
not give the exact entropy which requires us to evaluate the
appropriate Wald's integral at the horizon, the entropy calculated
by regarding the long throat region as the near horizon geometry
will continue to give an approximate value of the entropy. However
the entropy calculated this way  acquires a mild dependence on the
asymptotic moduli since the near horizon values of the originally
flat moduli depends on the asymptotic data, and the entropy function
now has a piece $\Delta\EE$ that depends on these `flat directions'.

Even though we have presented the problems with runaway behavior at
the horizon and that of defining statistical entropy of microscopic
states as two separate problems, we expect them to be related. In
the spirit of AdS/CFT correspondence we could identify the radial
evolution of various moduli fields in the black hole description
with the renormalization group (RG) evolution of various parameters
in the microscopic theory describing the black hole. Thus a runaway
behavior of the moduli fields in the gravity description will
correspond to a runaway behavior of the parameters of the
microscopic theory in the far infrared. Even if there is a
non-trivial infrared fixed point where the parameters reach a finite
value, either the gravity description, or the microscopic
description (or both) must be strongly coupled in this region since
we cannot have a configuration where both the gravity and the
microscopic description are simultaneously weakly coupled. The way
this problem is avoided in the case of supersymmetric black holes is
by having one or more flat directions of the near horizon geometry
which we can tune to go from weakly coupled microscopic description
to weakly coupled gravity description. Since for non-supersymmetric
black holes we expect the flat directions to be lifted in general,
the only way we can avoid this problem is by introducing a small
amount of non-extremality. On the black hole side it effectively
cuts off the  evolution of the moduli fields at certain radius. Its
counterpart on the microscopic side is to introduce certain infrared
cut-off. This is precisely the effect of introducing a small
temperature into the system. The long throat region with
approximately $\bf AdS_2\times S^n$ geometry on the black hole side
should correspond, on the microscopic side, to a range of scale
where all the $\beta$-functions are small and we have an
approximately conformal quantum mechanics.

This by itself of course does not solve the problem, since again if
the parameters in this throat region are such that the microscopic
theory is weakly coupled, then the gravity description has strong
curvature and vice versa. However often in this case we have one or
more approximate flat directions which we can adjust to go from
weakly coupled microscopic description to weakly coupled
supergravity description. Since the entropy function does not change
appreciably as we move along these flat directions we get a relation
between the statistical entropy and black hole entropy. However
since we now only have approximately flat directions, both entropies
acquire mild dependence on the energy cut-off and asymptotic moduli
in their subleading piece. As a result the comparison between the
weak coupling statistical entropy and the strong coupling black hole
entropy cannot be carried out to an arbitrary accuracy, but only up
to terms of a certain order which are not affected by the
ambiguities in the definition of the entropy introduced due to the
need of considering slightly non-extremal black holes. Clearly, the
relationship is most robust for the leading term for which all the
ambiguities mentioned above disappear.

There is however a potential danger with this argument. We have seen
that the black hole entropy function now acquires a piece
$\Delta\EE$ which gives subleading contribution of the entropy that
depends on the original `flat directions'. These contributions are
subleading as long as the effect of lifting of the flat direction is
a small effect. However in order to carry on our argument we need to
vary the asymptotic moduli all the way to the weak coupling regime
and this could push the near horizon field configuration to a regime
where the $\Delta\EE$ piece becomes large. If this happens then we
can no longer use our argument to show  the equality of macroscopic
and microscopic entropy.\footnote{In fact in the weak coupling
regime the statistical entropy of states within the mass range
$\epsilon$ as discussed above is equal to that computed in the free
theory, and hence is independent of the coupling constant. So the
issue really is whether the entropy acquires a non-trivial
dependence on the coupling constant in the transition region between
the weak and the `strong' coupling.} The entropy function formalism
by itself cannot tell us if this happens or not; we need to analyze
the dynamics on a case by case basis to settle this issue. The point
however is that often it is possible, based on other arguments, to
determine the order at which the equations of motion associated with
the flat directions begin receiving a non-trivial contribution. The
entropy function formalism then tells us that the dependence of the
entropy on the flat directions also begins at that order, -- the
point being that the function whose extremization gives the
equations of motion is the same function whose value at the extremum
gives the entropy. If the order at which this dependence begins
remains subleading even in the transition region between weak and
`strong' coupling, then our argument about the equality of
microscopic and macroscopic entropy remains valid. We shall
illustrate this in an explicit example in \S\ref{srunaway}.

There is one class of examples discussed in this paper which require
a slightly different treatment. These are the cases of small black
holes. In this case the microscopic theory is that of elementary
strings, and for weak coupling when we work within single string
Hilbert space, this theory is free (if we work in flat space) or
described by a sector of a 1+1 dimensional conformal field theory.
This makes the computation of the microscopic entropy easy. As a
consequence we should expect that the near horizon geometry of the
corresponding black hole cannot be described within supergravity
approximation. This is indeed true since the curvature at the
horizon is of the order of the string scale, and there is no flat
direction which we can adjust to change this. Nevertheless by
varying the asymptotic parameters (on which the entropy does not
depend as a consequence of the attractor mechanism) we can bring
 the solution  to
a form
 where certain scaling arguments apply;
and we can use them to determine the dependence
of the black hole entropy on the charges up to an overall numerical
constant even though the horizon geometry has strong
curvature\cite{Sen:1995in,Peet:1995pe,Sen:2005kj}. In the case
of four dimensional black holes it has been possible to
even compute the overall numerical
constant using various additional
techniques\cite{Dabholkar:2004yr,Dabholkar:2004dq,Kraus:2005vz,
Kraus:2005zm}. However since these computations require us to
go beyond supergravity approximation, there is no obvious
contradiction with the fact that the microscopic theory is weakly
coupled.

\section{Extremal Black Holes Without Supersymmetry}\label{blackcount}

In this section we will give several examples of extremal black
holes for which the weak coupling value of the statistical entropy
agrees with the `strong' coupling value of the black hole entropy.
We first discuss two simple examples in $\S{\ref{fivedcount}}$ and
$\S{\ref{fourdcount}}$ and then turn to more general black holes in
type-II and M-theory on Calabi-Yau spaces in $\S{\ref{general}}$.

\subsection{A Nonsupersymmetric black hole in five
dimensions}\label{fivedcount}

Let us
consider heterotic string theory
compactified on $\bf K \times S^1$ where $\bf K$ is either $\bf T^4$ or
$\bf K3$, resulting in a theory with sixteen or eight
supersymmetries.
We denote by
$x^m$ for $m = 6, 7, 8, 9$  the coordinates along $\bf K$, and by
$x^5$  the coordinate of the $\bf S^1$.

A basic example of a non-supersymmetric state in this theory is the
following. Consider a fundamental heterotic string winding state
wrapping $w$ times along $\bf S^1$ and carrying quantized momentum
$n$ along the same direction. Such a state satisfies the Virasoro
constraint $N_L -N_R = 1 + nw$, where $N_L$ is the oscillator number
of the left-movers and $N_R$ is the oscillator number of the right
movers.\footnote{In our convention the left-movers carry positive
momentum along $\bf S^1$. This differs from the convention of several
other papers in the literature where left-movers carry negative
momentum along $\bf S^1$.} When $n>0$ and large, this constraint is
satisfied by states which are in the right-moving ground state but
carry arbitrary left-moving oscillation. Since the supersymmetries
are carried by the right movers, this state, which we refer to as
the F1-P state, is BPS\cite{Dabholkar:1989jt}. On the other hand,
when $n < 0$ and large, this constraint is satisfied by a state in
the left-moving ground state but carrying arbitrary right-moving
oscillations. Such a state, which we refer to as the F1-\bP state is
no longer supersymmetric, and indeed breaks supersymmetry
completely. The F1-P state corresponds to a supersymmetric small
black hole and the F1-\bP  state corresponds to a non-supersymmetric
small black hole. We thus see that we can go from a supersymmetric
state to a non-supersymmetric state simply by flipping the sign of
the momentum. This is a consequence of the fact that the 1+1
dimensional world-sheet theory of the heterotic winding string is
chiral and only the right-movers carry supersymmetry.

In the type-I description, the heterotic fundamental string is dual
to the solitonic D1 brane \cite{Dabholkar:1995ep, Hull:1995nu,
Polchinski:1995df} which is also chiral. Because of the chirality,
the direction of the momentum along the soliton determines whether
the solution is supersymmetric or not. The D1-P state is
supersymmetric and the D1-\bP state is non-supersymmetric.

So far we have considered states which correspond to small black
holes, \emph{i.e.} black holes which have vanishing entropy in the
supergravity approximation. To get a state that corresponds to a
large black hole with finite area in supergravity, we add D5 branes
wrapped on ${\bf K\times S^1}$ and consider D1-D5-P or D1-D5-\bP
state. Let us denote the D1 and D5-brane charges and momentum along
$\bf S^1$  by $Q_1$, $Q_5$ and $n$ respectively. Here $Q_1$ and $Q_5$
are positive and $n$ can be positive or negative. Since  for $n<0$
supersymmetry is broken completely before adding the D5 branes, it
continues to be broken even after adding the D5 branes. The counting
of states for this configuration in the perturbative regime, where
the 't Hooft coupling of the gauge theory describing the low energy
dynamics of the brane system is small, can be performed as in
\cite{Dabholkar:1997rk}. The dominant contribution to the entropy
comes from the 1-5 strings, localized on the effective 1-brane along
the $x^5$ coordinate. There are $4Q_1 Q_5$ bosons and as many
Majorana
fermions coming from the bi-fundamentals from the 1-5 sector along
this effective brane. Thus the left as well as right-moving
central charge of the CFT describing
the dynamics of the effective string is $6Q_1Q_5$
and Cardy formula gives the resulting entropy
to be $2\pi\sqrt{Q_1 Q_5
|n|}$ for both signs of $n$.
On the other hand the black hole solution describing this
configuration is also easy to construct in the supergravity
approximation. One just takes the black hole solution describing the
D1-D5-P system or D1-D5-\bP system in the type IIB string
theory\cite{Strominger:1996sh} (both of which are supersymmetric)
and interprets it as a black hole solution in the type I theory
after the orientifold projection. From this it is clear that the
black hole will have the same entropy for either sign of $n$; indeed
the part of the low energy effective action of the type I string
theory that is relevant for describing this black hole solution  has
a $Z_2$ symmetry (that it inherits from the parent type IIB theory
and is broken once we take into account the effect of the
orientifold plane and the D9-branes) that allows us to relate the
black hole solutions for $n$ and $-n$. The answer for the black hole
entropy in the supergravity approximation is $2\pi\sqrt{Q_1 Q_5
|n|}$.

Thus we see that the statistical entropy based on weak coupling
counting agrees with the entropy of the corresponding black hole
which forms only when the 't Hooft coupling is large. We thus have
an agreement between the macroscopic and microscopic entropy even
though the states under consideration for $n<0$ break supersymmetry
completely and maximally.

\subsection{A nonsupersymmetric black hole
in four dimensions}\label{fourdcount}

In this section
we consider heterotic string theory
compactified on  $\bf K
\times S^1 \times {\wt S^1}$
where again $\bf K$ is either $\bf T^4$ or
$\bf K3$, resulting in a theory with sixteen or eight
supersymmetries.
We denote by
$x^m$ for $m = 6, 7, 8, 9$ the coordinates along $\bf K$, and by
$x^5$ and $x^4$  the coordinates of the $\bf S^1$ and
$\bf \wt
S^1$ respectively.

To obtain a four-charge large black hole in four dimensions we add
Kaluza-Klein 5-branes extending
along $56789$ directions
to the  configuration described in $\S{\ref{fivedcount}}$.
Since the type I D5-brane corresponds to heterotic 5-brane
lying along the 5-6-7-8-9 direction, in
the heterotic description we
have a configuration F1-NS5-KK5-P or
F1-NS5-KK5-\bP. Let $x^\mu$ be the
coordinates of the noncompact four dimensional
spacetime in which the
black hole
is located. The relevant vector potentials for describing the black
hole solution are $G_{4\mu}$ and
$G_{5\mu}$ coming from the metric and $B_{4\mu}$ and $B_{5\mu}$
coming from the 2-form. The  F1 and P (or \bP) are electrically
charged and couple to $G_{5\mu}$ and $B_{5\mu}$ respectively.
KK5 and NS5 are magnetically charged and couple to
$G_{4\mu}$ and $B_{4\mu}$ respectively. We always label the states
in this heterotic description and denote by
$Q_1$, $Q_5$, $\wt
Q_5$, and $n$  the numbers  of F1-strings, NS5-branes,
KK5-branes, and momentum in this duality frame.

The weak coupling counting is done most easily in the type-I$'$
description as  in \cite{Kaplan:1996ev}. First using the heterotic -
type I duality we map this system to a D1-D5-KK5-P or D1-D5-KK5-\bP
state in type-I theory. If we  now T-dualize the type-I theory along
the $x^4$ direction, then we obtain a D2-D6-NS5-P or D2-D6-NS5-\bP
state in type-I$'$.  The resulting configuration has $Q_5$ D6-branes
wrapping $456789$ directions, $Q_1$ D2-branes wrapping $45$
directions, $\wt Q_5$ NS5-branes wrapping $56789$ directions and
momentum $n$ flowing along the $5$ direction. In addition there are
O8-planes and D8-branes at the two ends of the $4$ direction. Since
D2-branes can end on an NS5-brane \cite{Strominger:1995ac}, the
presence of $\wt Q_5$ NS5-branes give rise to effectively $Q_1\wt
Q_5$ D2-branes. Therefore the microscopic entropy is given by
$2\pi\sqrt{Q_1Q_5\tQ_5|n|}$\cite{Maldacena:1996gb}. On the other
hand the black hole solution carrying these charges is identical to
a supersymmetric black hole solution carrying the same charges in
the parent type IIA theory before the orientifold projection, and
has an entropy $2\pi\sqrt{Q_1Q_5\tQ_5|n|}$ in the supergravity
approximation\cite{Maldacena:1996gb}. Thus the statistical entropy
is in agreement with the macroscopic black hole entropy. As we will
discuss in $\S{\ref{general}}$,  in the M-theory description
utilized in \cite{Maldacena:1997de} we can generalize this heuristic
counting to a larger class of black holes.

Even though we have considered a  four-charge system with a specific
charge assignment for simplicity of discussion, the conclusion can
be stated in a duality invariant way. Consider first the case of
heterotic on $\bf T^4 \times T^2$. The U-duality group in this case
is $O(6, 22, \bZ) \times SL(2, \bZ)$. Since we are dealing with
large black holes and supergravity action without higher derivative
corrections, we in fact have $O(6, 22, \bR) \times SL(2, \bR)$ at
our disposal. Let $Q$ be the electric charges and $P$ be the
magnetic charges of the black hole; they are both vectors of $O(6,
22, \bR)$. Then the black hole entropy in the supergravity
approximation can be written in a U-duality invariant way as $ S=
\pi \sqrt{|P^2 Q^2 - (P\cdot Q)^2|}$ where the dot product is the
$O(6, 22, \bR)$ invariant one. For the specific configuration
considered earlier we have $Q\cdot P=0$, $Q^2 = 2 Q_1 n$ and $P^2 =
2 Q_5 \wt{Q_5}$. Note in particular that for our non-supersymmetric
state $Q^2$ is negative since $n$ is negative. A supersymmetric
configuration on the other hand would have $Q^2$ positive. More
generally, for a general charge assignment the supersymmetric black
holes have the discriminant $P^2 Q^2 - (P\cdot Q)^2$ positive and
the non-supersymmetric black holes have $P^2 Q^2 - (P\cdot Q)^2$
negative in our conventions. The absolute value of the discriminant
is the one that enters into the expression for the entropy.

For the heterotic string on $\bf K3 \times T^2$, one obtains an
\CN=2 supergravity in four dimensions. The invariance of the
classical supergravity action in this case  is $O(2, n_v -1, \bR)
\times SL(2, \bR)$ where $n_v$ is the number of \CN=2 vector
multiplets. The formulae above apply with the only change that the
dot product is now the $O(2, n_v -1, \bR)$ invariant one.

\subsection{General extremal black Holes in M-theory on
$\bf CY_3\times S^1$}\label{general}

We will now consider a more general class of examples involving
black hole solutions in M-theory compactified on a circle $\bf S^1$
times a Calabi-Yau 3-fold $\bf CY_3$. By the usual duality between
type IIA   string theory and M-theory on $\bf S^1$, these can also
be regarded as black hole solutions in type IIA string theory on
$\bf CY_3$. We will consider the BPS black holes discussed in
\cite{Maldacena:1997de} with vanishing D6-brane charge but arbitrary
D4-brane charges $\{ p^A\}$, D2-brane charges $\{q_A\}$ and D0-brane
charge $q_0$.  Here the index $A = 1, 2, \ldots, n_v$ labels the
$n_v$ 4-cycles (or equivalently the dual 2-cycles) of $\bf CY_3$.
Thus we have $p^A$ D4-branes wrapped on the $A$-th 4-cycle
$\Sigma_A$, $q_A$ D2-branes wrapped on the $A$-th 2-cycle $\sigma^A$
and $q_0$ D0-branes. If we denote by $\bf P$ the four cycle $p^A
\Sigma_A$, then in the M-theory description this configuration
corresponds to a M5-brane wrapped on $\bf P\times S^1$, with
appropriate fluxes turned on the brane to produce the D2-brane
charges, and carrying $q_0$ units of momentum along $\bf S^1$. If
\textbf{P} is a `very ample' divisor, then it is smooth at a generic
point in the moduli space and an M5-brane wrapped on it is locally a
single, smooth brane. Its  massless fluctuation modes can then be
computed using index theory as in \cite{Maldacena:1997de} and is
summarized by a $(0, 4)$ superconformal field theory living on the
effective string wrapping the M-theory circle $\bf S^1$. The number
of massless left-moving and right-moving bosons and fermions on this
string deduced from index theory gives us the left-moving and
right-moving central charges $c_L$ and $c_R$ of the conformal field
theory (CFT).

Let us denote  by $N^B_L$ the left-moving bosons and by
$N^B_R$ and $N^F_R$  the right-moving bosons and fermions
respectively.    Also let $6D_{ABC} = \Sigma_A \bigcap
\Sigma_B \bigcap \Sigma_C$ be the intersection numbers of the
four cycles $\Sigma_A$.  The central charges are
then given by\cite{Maldacena:1997de}
\begin{eqnarray}\label{UV}
  c_L  &=& N^B_L = 6D + c_2 \cdot P \\
  c_R  &=& N^B_R + \half N^F_R = 6D + \half c_2 \cdot P,
\end{eqnarray}
where $D=D_{ABC}p^Ap^Bp^C$ and
$c_2 \cdot P \equiv c_{2A}p^A$, $c_{2A}$ being the second
Chern class of the four cycle $\Sigma_A$.
On the other hand the conformal
weight of the lowest energy state carrying the charges described above
is given by
\bea \label{econf}
(h_L, h_R) &=& (\hat{q_0}, 0)
\quad \hbox{for $\hat {q_0}>0$} \nonumber \\
&=& (0, -\hat {q_0}) \quad \hbox{for $\hat{q_0}<0$}\, ,
\eea
where
\begin{equation}\label{hat}
\hat{q_0} = q_0 + \frac{1}{12} D^{AB} q_A q_B,
\end{equation}
$D^{AB}$ being the inverse of $D_{AB} \equiv D_{ABC}p^C$.
Then according to Cardy formula the statistical entropy,
defined as the logarithm of the degeneracy of
states,  is given by
\bea \label{estat}
S_{stat} &=& 2\pi \sqrt{c_L h_L\over 6} = 2\pi \sqrt{(D +{1\over 6}
c_2\cdot P)\hat{q_0}}
\quad \hbox{for $\hat{q_0}>0$} \nonumber \\
&=& 2\pi \sqrt{c_R h_R\over 6} = 2\pi \sqrt{(D +{1\over 12}
c_2\cdot P)|\hat{q_0}|} \quad \hbox{for $\hat{q_0}<0$}\, .
\eea
The states with $\hat{q_0}>0$ are BPS,
whereas states with $\hat{q_0}<0$ break
all supersymmetries.\footnote{As explained in
\cite{Maldacena:1997de}, it is possible to maintain supersymmetry
even with  right-moving momentum as long as it is a multiple of
the integral class [P] in the momentum lattice. But a generic
right-moving momentum will break supersymmetry.}
Since $D$ is cubic in the charges $p^A$ whereas
$c_2\cdot P$ is linear in these charges, we have $D>> |c_2\cdot P|$.
Thus for both signs of $\hat{q_0}$ the leading contribution to the
statistical entropy is given by
\be\label{esstat}
S_{stat} = 2\pi \sqrt{D\, |\hat{q_0}|}\, .
\ee

The macroscopic entropy of the corresponding
black hole solution to leading order in
supergravity goes as\cite{Maldacena:1997de}
\begin{equation}\label{macromsw}
S_{BH} = 2\pi \sqrt{D|\hat{q_0}|},
\end{equation}
for both signs of $\hat {q_0}$. This approximation is valid for
large charges. Thus the statistical entropy \eq{estat} agrees with
the macroscopic entropy calculated in the supergravity approximation
in the large charge limit both for BPS as well as non-BPS states. In
fact in this case there are general arguments that this agreement
continues to hold for both BPS and non-BPS states even after
inclusion of higher derivative
corrections\cite{Kraus:2005vz,Kraus:2005zm,Sahoo:2006vz}. We will
return to this point in \S\ref{scomp}.

Other examples involving rotating black holes will be discussed in
\S\ref{srotate}.

\section{Geometry of the D1-D5-\bP System}\label{interpolating}

In this section we will  analyze in detail the five dimensional
black-hole with three charges discussed in $\S{\ref{fivedcount}}$.
For definiteness we will take the compact space to be $\bf T^4\times
S^1$, but the extension to the $\bf K3\times S^1$ case is
straightforward. In the type I description this state couples only
to the graviton $ G_{MN}$ and the dilaton $ \phi$ from the NS-NS
sector, and the 2-form potential $ B_{MN}$ from the R-R sector. The
low energy action for these fields is
\bea\label{tendaction}
S &=& \int d^{10} x \sqrt{-\det G} \LL,  \nonumber \\
\LL &=&
{1 \over {16 \pi G_{10}}}  \left[
                e^{-2 \phi}(R + 4\left(\nabla \phi)^2\right)
                 - {1 \over12} H^2 \right],
\eea
where $H$ is the 3-form field strength associated with $B_{MN}$
and
\be \label{enew1.5} 16\, \pi\,
G_{10}= (2\pi)^7 \,(\alpha')^4\, \ee would
be the ten dimensional Newton's constant if $\phi$ vanishes
asymptotically. The solution with three charges $Q_1$,  $ Q_5$, and
$ n$ with $n<0$ is the same as the corresponding solution in
type-IIB theory\cite{Dabholkar:1997rk}
\begin{eqnarray}\label{solution2}
\nonumber dS^2 &=& ( 1 + { r_1^2 \over r^2} )^{-1/2} ( 1 + { r_5^2
\over r^2} )^{-1/2}  [ - dt^2 +dx_5^2 + {r_n^2  \over r^2} (dt -
dx_5)^2
 +( 1 + { r_1^2  \over  r^2}) dx_i dx^i ] \\
&+& ( 1 + { r_1^2  \over r^2})^{1/2} ( 1 + { r_5^2 \over r^2}
)^{1/2} \left[ dr^2 + r^2 d \Omega_3^2 \right]
\end{eqnarray}
\begin{eqnarray}\label{solution1}
 H\equiv{1\over 6}H_{MNP}dx^M\wedge dx^N\wedge dx^P
 &=& 2 \lambda^{-1}\,
 r_5^2 \e_3 + 2 r_1^2 \lambda \, e^{-2 \phi}  *_6 \e_3,\nonumber \\
e^{-2\phi } &=& \lambda^{-2}\, (1+
{ r_5^2\over r^2 }) (1 + {r_1^2 \over  r^2
})^{-1},\quad
\end{eqnarray}
where
$x^5$ is the coordinate of a circle
$\bf S^1$ with coordinate radius $R$,
 $x^i$ for
$i = 6,...,9$ are the coordinates of a torus $\bf T^4$ with
coordinate volume ${\left( 2\pi \right)^4 V}$,
$\e_3$ is the volume element on the unit three-sphere
and $*_6$ denotes
the Hodge dual in the six dimensions spanned
by $x^0,..,x^5$.
Thus $ \lambda$, $(2\pi)^4V$, and $R$
are asymptotic values of the string
coupling constant, the volume of $\bf T^4$, and the radius of the
$\bf S^1$, all measured in string units.
This solution represents a black hole
in the five dimensional theory spanned by $x^0,\ldots x^4$.
The parameters of the solution
are related to the integral charges $Q_1$, $Q_5$ and $n$ through the
relations
\begin{equation}
r_1^2 = {\l Q_1 \apm \over V }, \qquad r_5^2 = {\l Q_5 \apm
},\qquad r_n^2 = { \l^2 |n |\apm \over R^2 V }\, .
\end{equation}
The term involving $(dt -
dx_5)^2$ in the metric (\ref{solution2}) corresponds to
right-moving momentum $n <0$ along the soliton and
the solution breaks all supersymmetries. If we instead use $n >0$
at asymptotic infinity then the solution depends on the
combination $(dt + dx_5)^2$ and supersymmetry is preserved.

It is instructive to study the near horizon geometry of this black
hole. For this we will set $\alpha'=1$ and define new coordinates
and parameters \bea \label{enew1} && \rho = r^2 / (r_n^2 R^2),
\qquad \tau = 2 r_n R^2 t / (r_1 r_5)
, \nonumber \\
&& y^5 = (x^5-t)/R, \qquad y^i = x^i/V^{1/4} \quad
\hbox{for}\quad 6\le i\le 9,
\eea
\bea \label{enew2}
&& v_1 = {r_1 r_5 \over 4}
={1\over 4} \, {\lambda\over \sqrt V}
\sqrt{Q_1 Q_5}, \qquad
v_2=r_1 r_5 = {\lambda\over \sqrt V}
\sqrt{Q_1 Q_5}, \nonumber \\
&&  u_1 = {r_n^2 R^2 \over r_1 r_5}
={\lambda\over \sqrt V } \, {|n|\over \sqrt{Q_1 Q_5}},
\qquad u_2= {r_1V^{1/2} \over r_5}
=\sqrt{Q_1\over Q_5},
\nonumber \\
&& u_3 = { r_5 \over r_1 \lambda}={\sqrt V\over \lambda}
\, \sqrt{Q_5\over Q_1}, \nonumber \\
&&
 e_1 =   {r_5 r_n R \over 4 \lambda r_1}={1\over 4}
 \sqrt{Q_5 |n|\over Q_1}, \qquad
 e_2 = -{r_1 r_5 \over 4 r_n R}
 =-{1\over 4} \sqrt{Q_1 Q_5\over |n|}\, ,
 \nonumber \\
\eea
and then take the $r\to 0$ limit.
With this definition $y^5$ has
coordinate radius 1, $y^6,\ldots y^9$ have coordinate
volume $(2\pi)^4$ and the near horizon geometry takes the form:
\bea \label{enew3}
dS^2 &=&   v_1 \left(-\rho^2 d\tau^2 + {d\rho^2\over
\rho^2}\right)
+ v_2 d\Omega_3^2
+ u_1 (dy^5- 2 e_2 \rho d\tau)^2 + u_2 dy^i dy^i \nonumber \\
 H &=& 2\, Q_5 \epsilon_3 + 2\,
 e_1 d\tau\wedge d\rho\wedge dy^5\, ,
\qquad e^{-2\phi} = u_3^2 \, .
\eea
{}From this we see that many of the fields in this geometry get
attracted to fixed values at the horizon. For example, the volume
of the $\bf T^4$ at the horizon gets attracted to $(2\pi)^4u_2^2=
(2\pi)^4Q_1 / Q_5$
independent of the asymptotic value $V$. Not all moduli get fixed,
however. For example, several parameters including the
dilaton at the horizon continue to
depend on the asymptotic modulus $V/\lambda^2$.
The entropy is, of course,
independent of all asymptotic
moduli and depends only on charges as
$2\pi\sqrt{Q_1 Q_5 \bar P}$.

We will now derive the near horizon geometry given in
eqs.\eq{enew2}, \eq{enew3} using the entropy function formalism
\cite{Ghodsi:2006cd}. For arbitrary parameters $v_1$, $v_2$, $u_1$,
$u_2$, $u_3$, $e_1$, $e_2$ and $p_1$,  eq.\eq{enew3} describes the
general background with zero Kaluza-Klein monopole charge associated
with the $y^5$ direction, preserving the $SO(2,1)\times SO(4)$
symmetry of $\bf{AdS_2\times S^3}$. In order to compute the entropy
function for this black hole we introduce normalized charges \be
\label{enew2.5} p^1 = Q_5, \qquad q_1 =   2 \, Q_1, \qquad q_2 =
2\, n\, , \ee and define\footnote{Since the dimensional reduction on
$\bf T^4\times S^1$ produces Chern-Simons terms, eq.(\ref{enew4}) is
not valid in general. One needs to first express the dimensionally
reduced Lagrangian density in a manifestly covariant form by
throwing away total derivative terms, and then define $f(\vec v,
\vec u, \vec e, \vec p)$ using this covariant Lagrangian density.
Typically this gives rise to additional contribution to $f$ besides
(\ref{enew4})\cite{Sahoo:2006pm}. However in the present example the
Chern-Simons terms in $H$ vanish and hence (\ref{enew4}) gives the
correct contribution to $f$.} \be\label{enew4} f(\vec v, \vec u,
\vec e, \vec p) = \int_H \sqrt{-\det G} \, \LL \ee evaluated in the
background \eq{enew3}. Here $\int_H$ denotes integration over the
horizon of the black hole. In the ten dimensional description this
is $\bf S^3\times S^1\times T^4$ labeled by the angular coordinates
labeling the three sphere, the coordinate $y^5$ and the coordinates
$y^6,\ldots y^9$. The entropy function $\EE (\vec v, \vec u, \vec e,
\vec q,\vec p)$ is then given by\cite{Sen:2005wa} \be\label{enew5}
\EE (\vec v, \vec u, \vec e, \vec q,\vec p) =2\pi\left( q_1 e_1 +
q_2 e_2 - f(\vec v, \vec u, \vec e, \vec p) \right)\, , \ee and the
entropy of the extremal black hole for a given set of electric
charges $(\vec q,\vec p)$ is obtained by extremizing the entropy
function with respect to the variables $v_i$, $u_i$ and $e_i$.

In the present problem the function $f$ can be easily evaluated
and is given by
\be \label{enew6}
f(\vec v, \vec u, \vec e, \vec p) = {4\pi^6\over G_{10}}\,
v_1 v_2^{3/2} \sqrt{u_1} u_2^2 \left[
u_3^2 (-2 v_1^{-1} + 6 v_2^{-1} + 2 u_1 v_1^{-2} e_2^2)
+ 2 u_1^{-1} v_1^{-2} e_1^2 - 2 v_2^{-3} p_1^2  \right]\, .
\ee
This gives, using $G_{10}=8\pi^6\, (\alpha')^4=8\pi^6$,
\bea\label{enew7}
\EE (\vec v, \vec u, \vec e, \vec q,\vec p) &=& 4\pi \Bigg[
Q_1 e_1 + n e_2
- {1\over 4} \,
v_1 v_2^{3/2} \sqrt{u_1} u_2^2 \bigg\{
u_3^2 (-2 v_1^{-1} + 6 v_2^{-1} + 2 u_1 v_1^{-2} e_2^2)\nonumber
\\
&&
+ 2 u_1^{-1} v_1^{-2} e_1^2 - 2 v_2^{-3} Q_5^2 \bigg\}
\Bigg]\, ,
\eea
where we have used \eq{enew1.5}
and replaced $q_1$, $q_2$ and $p_1$ in terms of $Q_1$,
$n$ and $Q_5$ using \eq{enew2.5}.
It is easy to see that this function has an extremum at
\bea \label{em1}
&& v_1
={1\over 4} \, \xi\,
\sqrt{Q_1 Q_5}, \qquad
v_2=\xi\,
\sqrt{Q_1 Q_5}, \qquad
u_1  =\xi \, {|n|\over \sqrt{Q_1 Q_5}},
\qquad u_2
=\sqrt{Q_1\over Q_5}, \nonumber \\
&&
 u_3 =  \xi^{-1}\,
\, \sqrt{Q_5\over Q_1},
\qquad e_1 =    {1\over 4}
 \sqrt{Q_5 |n|\over Q_1}, \qquad
 e_2 =  -{1\over 4} \sqrt{Q_1 Q_5\over |n|}\, ,
 \nonumber \\
\eea
for $n<0$.
Here $\xi$ is an arbitrary parameter reflecting a flat
direction of the entropy function. This agrees with \eq{enew5}
for $\xi=\lambda/\sqrt V$. Furthermore the value of $\EE $ evaluated
at this extremum is
\be \label{em1.5}
\EE  = 2\pi \sqrt{Q_1 Q_5 |n|}\, .
\ee
This reproduces the entropy of this black hole.

The same conclusions can also be reached using the effective
potential described in \cite{Ferrara:1997tw,Goldstein:2005hq}.  One
finds that the effective potential is extremized for the values of
the moduli given in eq.(\ref{em1}). The extremum has one flat
direction and is a
minimum along the two other directions in moduli
space. This shows that the extremum is an attractor along the two
non-flat directions.  For the supersymmetric case the attractor
behavior is expected.  In the non-supersymmetric case it follows
from an invariance of the effective potential under the charge
conjugation symmetry, $n \rightarrow -n$.

Now from eq.(\ref{em1}) we see that
as long as $Q_1$, $Q_5$ and $n$ are large, --
say $Q_1\sim N$, $Q_5\sim N$, $|n|\sim N^2$ with $N$ large.
 -- all scalars constructed out of curvature and gauge field
strengths at the horizon are small for finite $\xi$. Thus the
supergravity approximation is reliable. Furthermore, assuming that
the basic symmetry of the attractor geometry does not change from
$\bf AdS_2 \times S^{3}$, one can evaluate the entropy function of
\cite{Sen:2005wa,Sen:2005iz} to find that the higher derivative
terms give subleading corrections.  Since the attractor values of
the scalars are  determined by minimizing the entropy function and
the Bekenstein-Hawking-Wald entropy is the value of this function at
the minimum,  the resulting entropy will have a sensible
perturbative expansion in inverse powers of $N$. Furthermore, since
the entropy is independent of $\xi$, the answer \eq{em1.5} will
continue to be  valid even if $\xi$ is small. In particular, for the
scaling of $Q_1$, $Q_5$ and $|n|$ given above if we want the
effective interaction strength in the microscopic theory to be
small. we need to take $\xi N$ to be a small number. In this region
$v_1$, $v_2$ and $u_1$ are small indicating that the higher
derivative corrections become important. Nevertheless our argument
shows that the Wald entropy will continue to be given by
(\ref{em1.5}).

The argument given above assumes that the flat direction labeled by
$\xi$ is not lifted when we add higher derivative terms to the
action. For  supersymmetric black holes, -- {\it e.g.}  the one
obtained by replacing $n\to -n$, $e_2\to -e_2$ in the solution
described above, -- we expect this to be true. As a result the value
of $\xi$ at the horizon is a free parameter and the value of the
entropy is independent of this parameter. However for the
non-supersymmetric black holes the flat directions may get lifted
under addition of higher derivative corrections at some
order.\footnote{In the heterotic description the parameter
$\xi^{-2}$ actually correspond to the volume of the $\bf T^4$
measured in the string metric. Since there are no charges associated
with the gauge field arising out of $\bf T^4$ compactification, the
full black hole geometry is a product space of $\bf T^4$ and a six
dimensional manifold labeled by $x^0,\ldots, x^5$. This explains why
in the supergravity approximation the modulus $\xi$ does not get
fixed by the attractor mechanism. In fact this feature continues to
hold even after inclusion of tree level higher derivative
corrections in the heterotic string theory. The loop corrections
however will couple the black hole geometry and the six dimensional
geometry and is expected to generate a potential for $\xi$.} In that
case the parameter $\xi$ appearing in \eq{em1} will no longer be
arbitrary and will take some fixed value independent of the
asymptotic moduli. As long as $\xi N$ at the fixed point is large
the horizon geometry has low curvature and higher derivative
corrections are small. However if $\xi N$ becomes of order one or
less, we have highly curved horizon geometry and the derivative
expansion is no longer sensible for the computation of the entropy
function.\footnote{Note that since now $\xi$ at the horizon is
independent of the asymptotic coupling, this problem exists even
when the asymptotic 't Hooft coupling is large.} In \S\ref{srunaway}
we will give a uniform treatment of all these cases by introducing a
small amount of non-extremality to control the effect of non-trivial
dependence of the entropy function on $\xi$.

So far most of our attention has been focussed on the near horizon
geometry. Let us now look closely at the full interpolating geometry
given in (\ref{solution2}), (\ref{solution1}). First consider the
case $\lambda N>>1$. In this case $r_1$, $r_5$ and $r_n$ are large.
The near horizon geometry $\bf AdS_2 \times S^{3}$ is obtained if we
can ``drop the one'' in the harmonic functions appearing in the
equations (\ref{solution2}) and (\ref{solution1}). This can be done
once $r << r_1, r_5, r_n$ simultaneously.  Since $r_1$, $r_5$ and
$r_n$ are large we can ``drop the one'' even if $r$ remains large
compared to the string scale and one never runs into a high
curvature region in the interpolating geometry all the way from the
asymptotic infinity to the horizon. In this regime, higher
derivative corrections to the solution are small throughout the
entire geometry. Now consider what happens when we start reducing
the asymptotic coupling $\lambda$ keeping $N$ fixed at some large
value. Once $\lambda N$ becomes of order 1, the radii $r_1$, $r_5$
are no longer large and in order to reach the near horizon geometry
we need to take $r<< 1$. Thus the geometry enters  the large
curvature region $r\sim 1$. In this region corrections to the action
due to higher derivative terms are no longer small and we do not
have a systematic approximation scheme for calculating these
corrections. Nevertheless, as long as the solution approaches the
${\bf AdS_2\times S^3}$ form given in \eq{enew3} for small $r$, the
near horizon geometry and entropy are determined by extremizing the
entropy function and as a result the entropy is equal to its value
for $\lambda N>>1$ as long as we can ignore the issue of lifting of
the flat direction.

\section{Taming the Runaway}\label{srunaway}

In this section we will address the potential problem with the
runaway behavior of near horizon parameters after inclusion of
higher derivative corrections to the supergravity action. In
particular, we are interested in a situation where the leading
two-derivative action gives rise to a flat direction of the entropy
function or equivalently the effective potential.  In such a case,
the higher derivative corrections to the entropy function could lift
the flat directions in such a way that the entropy function has no
extremum. This would result in runaway behavior. What is the meaning
of the entropy calculated in the leading two-derivative
approximation in such a situation? In answering this question it is
useful to regard the entropy of the extremal black hole as a limit
of the entropy of a non-extremal black hole. By taking a slightly
non-extremal black hole, and large enough charge, we will see below
that the run-away behavior is in effect ``cut-off''. Since the black
hole is only slightly non-extremal the entropy would be close to
that of the extremal case calculated in the two-derivative
approximation.

Even though we will discuss the issue in the context of four
dimensional examples, the analysis easily generalizes to other
dimensions. We use the notation of \cite{Goldstein:2005hq} and
consider a theory with a lagrangian density of the form
\bea\label{ell1} \sqrt{-\det g}\, \LL &=&  {1\over \kappa^2} \left[R
- 2\, g^{\mu\nu} \p_\mu \phi_i \, \p_\nu \phi_i - f_{ab}(\vec \phi)
F^a_{\mu\nu}
F^{b\mu\nu} \right. \nonumber \\
&& \left. -{1\over 2} \, (\sqrt{-\det g})^{-1}\, \wt f_{ab}(\vec
\phi) \epsilon^{\mu\nu\rho\sigma} \, F^a_{\mu\nu} \,
F^b_{\rho\sigma} \right]\, , \eea where $g_{\mu\nu}$ denotes the
metric, $\{\phi_i\}$ denote a set of neutral scalar fields and
$F^a_{\mu\nu}=\p_\mu A^a_\nu -\p_\nu A^a_\mu$ denote a set of gauge
field strengths. $f_{ab}(\phi)$ and $\wt f_{ab}(\vec\phi)$ are a set
of functions which are fixed for a given theory. In this theory we
look for a spherically symmetric black hole solution of the
form:\footnote{In (\ref{ell2}) we have fixed the form of the gauge
field strengths by requiring that they solve the Bianchi identities
and field equations.} \bea\label{ell2} ds^2&\equiv& g_{\mu\nu}
dx^\mu dx^\nu = -a(r)^2 dt^2 + a(r)^{-2} dr^2 + b(r)^2 (d\theta^2 +
\sin^2\theta d\phi^2)\, , \cr {1\over 2}\, F^a_{\mu\nu} dx^\mu
\wedge dx^\nu &=& Q_{(m)}^a \sin\theta d\theta \wedge d\phi
+f^{ab}(\vec\phi) \, (Q_{(e)b} - \wt f_{bc}(\vec\phi) Q^c_{(m)}) \,
b(r)^{-2}\, dt\wedge dr \cr \phi_i &=& \phi_i(r) \, , \eea where
$f^{ab}(\vec\phi)$ is the matrix inverse of $f_{ab}(\vec\phi)$,
$Q^c_{(m)}$ and $Q_{(e)c}$ denote respectively the magnetic and
electric charges associated with the gauge field $A^c_\mu$, and
$a(r)$, $b(r)$ and $\phi_i(r)$ are functions to be determined. The
equations determining the radial evolution of $a(r)$, $b(r)$ and
$\phi_i(r)$ can be derived from a one dimensional
lagrangian\cite{Goldstein:2005hq} \be\label{ell3} {2\over \kappa^2}
\int \, dr\, \left[(a^2 b)' \, b' - a^2 b^2 (\phi')^2 -b^{-2}
V_{eff}(\vec\phi) \right] \ee where prime denotes derivative with
respect to the radial variable $r$ and \be\label{ell4}
V_{eff}(\vec\phi) = f^{ab}(Q_{(e)a}-{\wt f}_{ac}
Q^c_{(m)})(Q_{(e)b}- {\wt f}_{bd}Q^d
_{(m)})+f_{ab}Q^a_{(m)}Q^b_{(m)}\, . \ee If $V_{eff}(\vec\phi)$ has
a minimum at $\vec \phi=\vec\phi_0$ with $V_{eff}(\vec\phi_0)=Q^2$,
and if we parametrize the radial coordinate $r$ in such a way that
the horizon is at $r=Q$, then for an extremal black holes, as $r\to
Q$ we have\cite{Goldstein:2005hq} \be\label{elimit} \vec\phi(\vec
r)\to \vec\phi_0, \qquad a(r)\to {r-Q\over Q}, \qquad b(r)\to Q\, .
\ee This describes an $\bf AdS_2\times S^2$ near horizon geometry.

The effective potential $V_{eff}(\vec\phi)$
typically has some flat directions and hence
a family of minima. At
the minima
some moduli $\chi_\alpha$ are fixed to be $\chi_{\alpha}^*$, but
some moduli, representing deformations
along these flat directions, are not fixed.
For simplicity we consider the case where there is
only one such flat direction and label the coordinate along this
direction by $\xi$.
An example
of such a flat direction
is provided by the case
discussed in \S\ref{interpolating}, where the flat
direction is also called $\xi$.

For simplicity, in the analysis below
we will set the asymptotic values of all the moduli $\chi_\alpha$ to
their attractor values $\chi_{\alpha}^*$,
so that in the leading supergravity
approximation these moduli remain constant for all $r$:
$\chi_\alpha(r)=\chi_{\alpha}^*$.
In this approximation the $\xi$ modulus is
also independent of $r$ since the effective potential is $\xi$-independent.
With these boundary conditions,
the leading effective potential evaluated on the solution is a constant,
independent of $r$.
It is also independent of the flat direction  $\xi$.
So we write
\be\label{ell5}
V_{eff}|_{\hbox{solution}}=Q^2\, ,
\ee
where $Q$ is a constant independent of $r$.
The resulting solution is the extremal Reissner-Nordstrom
black hole,
\be\label{ell6}
a^2=(1-Q/r)^2 \qquad b=r\, .
\ee
Note that in our conventions the parameter
$Q$ has dimension of length.
There are also non-extremal black holes. These have,
\be
\label{ne}
a^2=(1-{\alpha \over r}) (1-{\beta\over r}), \quad b=r,
\ee
with
\be\label{ell7}
\alpha \beta =Q^2\, .
\ee
We take $\alpha>\beta$
by convention, so that the outer horizon is at $r=\alpha$.
A slightly non-extremal black hole has,
\be \label{ell8}
{(\alpha -\beta) \over \alpha} \ll 1\, .
\ee

Let us now ask what happens when
higher derivative terms contribute an extra term $h(\xi)$ in $V_{eff}$
so that $V_{eff}(\vec\phi)$ has the form
\be
\label{veffa}
V_{eff}(\vec\phi)
= f^{ab}(Q_{(e)a}-{\wt f}_{ac}
Q^c_{(m)})(Q_{(e)b}- {\wt f}_{bd}Q^d
_{(m)})+f_{ab}Q^a_{(m)}Q^b_{(m)}
+ h(\xi)\, .
\ee
The resulting one dimensional action is then, \be
\label{actob} S={2\over \kappa^2}\int dr\left((a^2b)'b'-a^2b^2
(\xi')^2 -{V_{eff} \over b^2} \right), \ee with $V_{eff}$ given in
eq.(\ref{veffa}).
$h(\xi)$, having its origin in four and higher derivative terms
in the effective action, is of order $Q^{-k}$ with $k\ge0$ for
$\xi\sim 1$.\footnote{We could include a dependence of $h$ on the
other moduli fields $\{\chi_\alpha\}$, but this
will not affect our main conclusions.
Also, strictly speaking if the additional terms
are arising due to higher derivative
corrections,  we need to keep other higher
 derivative terms in the
analysis, for example, in the kinetic energy terms for scalars etc.
In general after inclusion of these terms the equations of motion
will have more solutions some of which could diverge at the horizon.
We are assuming that if we choose the solution that is regular at
the horizon then it can be matched on to the asymptotically flat
Minkowski space-time.  We expect that for such solutions the effect
of these higher derivative terms will remain small all through the
solution and will not change our main conclusion that for big enough
$Q^2$, if the black hole is only slightly non-extremal,  $\xi$
essentially does not evolve from its value at $\infty$ all the way
to the horizon.}.

We will consider a non-extremal black hole and will self
consistently solve the equations by assuming that $\xi$ does not
vary significantly from its asymptotic value at $r=\infty$ all the
way till the horizon of the black hole.\footnote{It is not necessary
to consider the evolution all the way from $\infty$ to the horizon.
In particular when the asymptotic coupling constant is small, we
expect that the curvature and other field strengths will become
large in an intermediate region where the higher derivative terms
play an important role. Nevertheless the geometry is expected to
emerge into an approximately $\bf AdS_2\times S^2$ geometry sufficiently
close to the horizon. We can then concentrate on the radial
evolution of $\xi$ in this region, and show that $\xi$ does not vary
appreciably in this region.} Consistent with this assumption, to
leading order $\xi(r)=\xi(\infty)$. As long as $\xi(\infty)$ is of
order one, the effective potential is approximately
$$V_{eff}\simeq Q^2+h(\xi(\infty)).$$
To this order the metric of a slightly non-extremal black hole is
then given by  eq.(\ref{ne}) with \be\label{eabq}
\alpha\beta=Q^2+h(\xi(\infty))\, . \ee

We now turn to calculating the radial evolution of $\xi$.
Since the only $\xi$ dependence of $V_{eff}(\vec\phi)$ comes
from the $h(\xi)$ term in (\ref{veffa}), $\xi$
satisfies the equation,
\be
\label{eqxi}
\partial_r(a^2b^2 \partial_r\xi)={g(\xi) \over 2 b^2}
\ee where \be\label{egxi} g(\xi) = \p_\xi h(\xi)\, . \ee In writing
down (\ref{eqxi}) we have assumed that $\xi$ is a canonically
normalized field. To calculate the first corrections we will set
$\xi=\xi(\infty)$ on the right hand side of (\ref{eqxi}) and then
solve this equation. This gives, \be \label{valxi}
\xi(r)={g(\xi(\infty)) \over 2 \alpha \beta}\ln( {r-\beta \over r})
+ \xi(\infty)\, . \ee In arriving at (\ref{valxi}) we have fixed an
integration constant so that the solution is non-singular at the
horizon $r=\alpha$. Indeed, from (\ref{valxi}) we see that $\xi(r)$
approaches a finite limit as we approach the horizon $r=\alpha$. If
however we take the extremal limit when $\alpha=\beta$, $\xi(r)$ has
a runaway behavior as we approach the horizon unless
$g(\xi(\infty))=0$. In the full solution this condition takes the
form $g(\xi(\alpha))=0$, \emph{i.e.} $\xi$ should approach an
extremum of $h(\xi)$ as we approach the horizon. If $h(\xi)$ does
not have an extremum then there is no way to avoid the runaway
behavior.

Let us now return to the case of a near extremal black hole. For our
approximation to be self consistent, we need $\xi(\alpha) \simeq
\xi(\infty)$. More generally we require $\xi(r)$ in the whole range
between $\alpha$ and $\infty$ to be close to $\xi(\infty)$. This
means, from eq.(\ref{valxi}), \be \label{condc} \left|\ln({\alpha
-\beta \over \alpha})\right| \ll \left| {2 \alpha \beta
\over g(\xi(\infty))}\right|\, . \ee Using the leading order result
(\ref{ell7}) $\alpha \beta $ on the right hand side of
eq.(\ref{condc}) can be approximated by  $Q^2$. Using
eqs.(\ref{ell8}), (\ref{condc}) we now get \be \label{condf} 1\gg
{\alpha -\beta \over \alpha} \gg e^{-\left|{2  Q^2 \over
g(\xi(\infty))} \right|} \ee As long as
$g(\xi(\infty))\sim 1$, the term on the right hand side of
(\ref{condf}) is exponentially suppressed for large $Q$. Thus the
condition eq.(\ref{condf}) can be easily met by appropriate choice
of the non-extremality parameter. When this condition is met, the
entropy of the non-extremal black hole is approximately given by,
\be \label{enta} S_{BH}\simeq\pi Q^2 \ee which is the entropy of the
extremal black hole in the leading approximation. However since
$V_{eff}$ receives correction proportional to $h(\xi(\infty))$, we
expect that the entropy also receives a similar correction. Since
this clearly depends on the asymptotic value $\xi(\infty)$ of the
field $\xi$, we see that for non-extremal black holes of this type,
the attractor behavior breaks down at the order in which the
potential for $\xi$ is generated.

We note in passing that in any case the
entropy of an extremal black hole should be defined
by extrapolating the answer from
the non-extremal case down to the extremal case, since
sufficiently close to
extremality the thermal description breaks down
and a direct analysis based on thermodynamics becomes
unreliable\cite{Preskill:1991tb}.
For the thermal description to work, we need that
$(\partial T / \partial M) \ll 1 $
where $M$ is the mass of the black hole.
This gives rise to the condition
$
(\alpha -\beta) / \alpha \gg {l_{pl}^2/Q^2}$.
This is a stronger restriction than (\ref{condf})
when $Q$ is large.
Thus as the non-extremality
parameter $(\alpha-\beta)/\alpha$
is reduced,  the thermal description will break down
 before any appreciable running of $\xi$ field
 can occur outside the horizon.
Since the usual Bekenstein-Hawking entropy (and presumably its Wald
generalization) of the extremal black hole is obtained by
extrapolating the answer obtained at the stage when the thermal
description is still reliable, we see that the running of the
modulus $\xi$ plays no appreciable role if the entropy is obtained
using this procedure.

In summary, for a black hole  which is close but not very close to
extremality, one finds  that the modulus $\xi$ does not evolve an
appreciable amount outside the horizon. The entropy of the resulting
black hole is close to that of the extremal one obtained by keeping
the leading term in the effective potential as long as $\xi$ is of
order 1. This however is not the end of the story. In order to argue
that the black hole entropy remains approximately constant up to the
region of parameter space where the microscopic description is good,
we may need to continue the parameter $\xi$ into a region where the
near horizon geometry develops large curvature and hence the
function $h(\xi)$  becomes comparable to or larger than the leading
term. Can we argue that this does not happen? As discussed earlier,
one needs to address this question on a case by case basis.  We will
illustrate this in the context of the example described in
\S\ref{interpolating}. For $Q_1\sim N$, $Q_5\sim N$ and $|n|\sim
N^2$ with $N$ large, the microscopic description is good when
$\lambda\, N<<1$. For $V\sim 1$ this requires $\xi\, N<<1$.
Examining the near horizon geometry  given in (\ref{em1}) we see
that in this region the sizes of $AdS_2$ and $S^2$ become small, and
hence $\alpha'$ corrections in the type I description become
important. On the other hand the type I string coupling constant is
exceedingly small and hence we can ignore the loop corrections. Thus
the question is: do the $\alpha'$ corrections generate a
contribution to $h(\xi)$? To answer this question note that the
$\alpha'$ corrections in type I theory are the same as those in the
parent type IIB theory before the orientifold projection. Since the
corresponding black hole in the parent type IIB theory is
supersymmetric, we expect that in this case the near horizon value
of $\xi$ is arbitrary. Thus the same will hold true for the
$\alpha'$ corrected type I theory. This in turn shows that $h(\xi)$
does not receive any contribution due to $\alpha'$ correction.

Finally we note that even in situations where $\xi$ is not a runaway
direction and  the full entropy function does have an extremum as a
function of $\xi$, we can still regulate the evolution of $\xi$ in
the $\bf AdS_2$ throat using the trick described in this
section.\footnote{In fact, we may be forced to do this to make the
computation of statistical entropy well defined.} By introducing a
small non-extremality parameter we can ensure that $\xi$ at the
horizon does not change by an appreciable amount from its asymptotic
value. The entropy of such black holes remain close to the one found
in the leading approximation, and hence can be computed reliably
using the entropy function method.

\section{Comparison Between $\bf AdS_2$ and $\bf AdS_3$ Based Approaches \label{scomp}}

For some extremal black holes in string theory the $\bf AdS_2$
component of the near horizon geometry, together with an internal
circle, describes a locally $\bf AdS_3$ space. More specifically the
near horizon geometry of these extremal black holes correspond to
that of extremal BTZ black holes\cite{Banados:1992wn} in $\bf AdS_3$
with the momentum along the internal circle representing the angular
momentum of the black hole\cite{Strominger:1997eq}. In such
situations, alternative arguments are available for  explaining the
agreement between the leading order thermodynamic and statistical
entropy. These arguments are quite powerful and applicable even for
non-BPS extremal black holes. In particular the enhanced isometry
group of the $\bf AdS_3$ space allows us to get a more detailed
information about the entropy of the system and prove certain
non-renormalization
theorems\cite{Kraus:2005zm,Kraus:2005vz,Kraus:2006wn} for the
entropy of supersymmetric as well as non-supersymmetric black holes.
In this section we will outline these arguments both from
macroscopic and microscopic points of view  so as to clearly
distinguish them from the more general argument presented in this
paper, and also carry out a comparison between the two approaches
when both methods are available.

The rest of this section is organized as follows. In \S\ref{scomp1}
we review the computation of the macroscopic entropy based on the
$\bf AdS_3$ near horizon geometry and compare the relative strength
and weaknesses of the $\bf AdS_3$ and $\bf AdS_2$ based approaches.
In \S\ref{scomp3} we give examples of extremal BPS and non-BPS black
holes in string theory which do not have any $\bf AdS_3$ factor so
that the arguments of \cite{Kraus:2005zm,Kraus:2005vz,Kraus:2006wn}
cannot be applied directly on such black holes. In \S\ref{scomp2} we
will discuss the microscopic description of  black holes  with
locally $\bf AdS_3$ near horizon geometry and its implication for
the non-renormalization of the statistical entropy of the system.

\subsection{Black holes with $\bf AdS_3$ near horizon geometry \label{scomp1}}

We begin by reviewing the origin of the $\bf AdS_3$ geometry. For
this we focus on the $\bf AdS_2$ part of the near horizon geometry
together with the electric flux through it. By choosing the basis of
gauge fields appropriately we can arrange that only one gauge field
has non-vanishing electric flux through the $\bf AdS_2$; let us
denote this gauge field strength by $F_{\mu\nu}=\p_\mu A_\nu -\p_\nu
A_\mu$. Then the relevant part of the near horizon background takes
the form: \be \label{ecomp1} ds^2 \equiv g_{\alpha\beta}dx^\alpha
dx^\beta = v_1 (-r^2 dt^2 + r^{-2} dr^2), \quad F_{rt} = e\, . \ee
Let us now assume that there is an appropriate duality frame in
which we can regard the gauge field component $A_\mu$ as coming from
the component of a three dimensional metric along certain internal
circle. Let  $\phi$ be the scalar field representing the metric
component along the extra circle. Then the three dimensional metric
can be expressed in terms of the two dimensional fields as
\be\label{ecomp2} ds_3^2 = \phi \left(g_{\alpha\beta} dx^\alpha
dx^\beta + (dy + A_\alpha dx^\alpha)^2 \right) \ee where $y$ denotes
the coordinate along the circle. For definiteness we will assume
that $y$ has period $2\pi$. The relation between the three
dimensional metric and the two dimensional metric given above is
somewhat non-standard;  this is related to the standard form by a
rescaling of the two dimensional metric by $\phi$. Since
(\ref{ecomp1}) gives $A_t=er$, and since the scalar field $\phi$
must take some constant value $u$ at the horizon, we see that the
three dimensional near horizon metric has the form \be
\label{ecomp3} ds_3^2 = u \left[ v_1 (- r^2 dt^2 + r^{-2} dr^2) +
(dy + erdt)^2 \right]\, . \ee One can show that if $v_1$ and $e$
satisfy the relation \be\label{ecomp4} v_1 = e^2\, , \ee then the
three dimensional metric (\ref{ecomp3}) describes a locally $\bf
AdS_3$ space. Had the coordinate $y$ taken values along a real line,
it would be a globally $\bf AdS_3$ space; however because of the
periodic identification we have a quotient of the $\bf AdS_3$ space
by a translation by $2\pi$ along $y$. The effect of taking this
quotient is to break the $SO(2,2)$ isometry group of $\bf AdS_3$ to
$SO(2,1)\times U(1)$, -- the symmetries of an $\bf AdS_2\times S^1$
manifold. Since the physical radius of the $y$ circle is given by $
\sqrt{G_{yy}}=\sqrt{u}$, we expect that the effect of this symmetry
breaking will be small for large $u$.

Let us for the time being ignore the effect of this symmetry
breaking and suppose that the background has full symmetries of the
$\bf AdS_3$ space. In this case we expect that the dynamics of the
theory in this background will be governed by an effective three
dimensional action, obtained by treating all the other directions,
including the azimuthal and polar coordinates $\phi$ and $\theta$
labeling the non-compact part of space, as compact. This effective
action will have the form \be \label{ecomp6} \int d^3 x \,
\sqrt{-\det G} \, (\LL_0^{(3)} + \LL_1^{(3)})\, , \ee where
$\LL_0^{(3)}$ is a lagrangian density with manifest general
coordinate invariance, and $\sqrt{-\det G}\, \LL_1^{(3)}$ denotes
the gravitational Chern-Simons term: \be \label{ecomp7} \sqrt{-\det
G}\, \LL_1^{(3)} = K\, \Omega_3\, , \ee $\Omega_3$ being the Lorentz
Chern-Simons 3-form and $K$ is a constant. One can then show, both
in the Euclidean action formalism
\cite{Kraus:2005vz,Kraus:2005zm,Solodukhin:2005ah} as well as using
Wald's formula \cite{Saida:1999ec,Sahoo:2006vz}, that the entropy of
the black hole with near horizon geometry described in
(\ref{ecomp3}) has  the form: \bea \label{ecomp8} S_{BH} &&= 2\pi
\sqrt{ c_L \, n\over 6} \quad \hbox{for} \, n>0\, ,
\nonumber \\
&&= 2\pi \sqrt{c_R \, |n|\over 6} \quad \hbox{for} \, n<0\, , \eea
where $n$ is the electric charge associated with the gauge field
$A_\mu$, and \be \label{ecomp9} c_L = 24\, (-g(l)+\pi\, K)\, ,
\qquad c_R = 24\, \pi\, (-g(l)-\pi\, K) \, , \ee \be\label{ecomp10}
g(l) = {1\over 4}\, \pi \, l^3 \, \LL_0^{(3)}, \qquad l =
2\sqrt{ue^2} \, . \ee $\LL_0^{(3)}$ in (\ref{ecomp10}) has to be
evaluated on the near horizon background (\ref{ecomp3}). This gives
a concrete form of the $n$ dependence of the entropy in terms of the
constants $c_L$ and $c_R$.

The constants $c_L$ and $c_R$ given in (\ref{ecomp9})
can be interpreted
as the left- and right-moving central charges of the two dimensional
CFT living on the boundary of the
$\bf AdS_3$\cite{Kraus:2005vz,Kraus:2005zm,Solodukhin:2005ah}.
$|n|$ has the interpretation of $L_0$ (or $\bar L_0$) eigenvalue
of the state in this CFT,
and (\ref{ecomp8}) can be interpreted as the Cardy formula in this
CFT.
This observation by itself does not give any further information
about the values of $c_L$ and $c_R$, but
a further simplification occurs if the theory has sufficient
number of supersymmetries. If
the boundary theory happens to have $(0,4)$ supersymmetry, then
the central charge $c_R$ is related to the central charge of an
$SU(2)_R$ current algebra which is also a part of the $(0,4)$
supersymmetry algebra. Associated with the $SU(2)_R$ currents there
will be $SU(2)$ gauge fields in the bulk, and the central charge of
the $SU(2)_R$ current algebra will be determined in terms of the
coefficient of the gauge Chern-Simons term in the bulk theory. This
determines $c_R$ in terms of the coefficient of the gauge Chern-Simons
term in the bulk theory\cite{Kraus:2005vz,Kraus:2005zm}.
On the other hand from
(\ref{ecomp9}) we see that
 $c_L-c_R$ is determined in terms of the
coefficient $K$ of the
gravitational Chern-Simons term.
Since both $c_L$ and $c_R$ are determined in terms of the
coefficients of the Chern-Simons term in the bulk theory, they do
not receive any higher derivative corrections. This completely
determines the entropy from (\ref{ecomp8}). Furthermore the
expression for the entropy derived this way is independent of all
the near horizon parameters and hence also of the asymptotic values
of all the scalar fields. Thus the entropy remains unchanged as we
go from the `strong' coupling regime to the weak coupling regime.

Clearly the existence of an $\bf AdS_3$ factor in the near horizon
geometry gives us results which are much stronger than the ones
which can be derived based on the existence of only an $\bf AdS_2$
factor. However, as indicated above, these results are valid only if
the physical radius of the compact $y$ coordinate is large.
Typically near horizon value of the radius of the $y$ direction is
fixed by the  entropy function extremization conditions (the
attractor equations) and is not a free parameter. If the charges
carried by the black hole are large but all of the same order then
the sizes of the compact directions are also of order unity. In this
case we expect the SO(2,2) symmetry of $\bf AdS_3$ to be broken
strongly. As a result the effective two dimensional action governing
the dynamics in $\bf AdS_2$ space, besides having a `local' piece of the
form (\ref{ecomp6}), contains additional terms which cannot be
written as dimensional reduction of a generally covariant three
dimensional action. There are various sources of these additional
terms, {\it e.g.} due to the quantization of the momenta along the
$y$ direction, contribution to the effective action from various
euclidean branes wrapping the $y$ circle, etc. In the presence of
such terms there will be additional contribution to the entropy
which are not of the form (\ref{ecomp8}). These additional
corrections can be interpreted as due to the corrections to the full
string theory partition function on thermal
$\bf AdS_3$\cite{Kraus:2006nb,Kraus:2006wn} or equivalently as
corrections to the Cardy formula in the CFT living on the boundary
of $\bf AdS_3$, but there is no simple way to calculate these
corrections without knowing the details of this CFT.

We will illustrate this by an example. We consider heterotic string
theory compactified on $\bf T^4\times S^1\times \wt S^1$ and
consider an extremal dyonic black hole in this theory with $n$ units
of momentum and $w$ units of fundamental string winding along $\bf
S^1$ and $\wt N$ units of Kaluza-Klein monopole charge and $\wt W$
units of H-monopole charge along $\bf \wt S^1$. In the leading
supergravity approximation the near horizon values of the radii $R$
and $\wt R$ of $\bf S^1$ and $\bf \wt S^1$ and field $S$
representing square of the inverse string coupling are given by (see
{\it e.g.} \cite{Sen:2005iz}) \be\label{ecomp11} R =
\sqrt{\left|n\over w\right|}, \qquad \wt R = \sqrt{\left| \wt W\over
\wt N\right|}, \qquad S = \sqrt{\left| {nw\over \wt N \wt
W}\right|}\, . \ee Furthermore the entropy is given by
\be\label{ecomp12} S_{BH} = 2\pi\, \sqrt{\left|nw\wt N\wt
W\right|}\, . \ee This clearly has the form given in (\ref{ecomp8})
with $c_L=c_R= 6|w \wt N\wt W|$. This is a consequence of the fact
that the circle $\bf S^1$ and the near horizon $\bf AdS_2$ geometry
combines into an $\bf AdS_3$ space if we treat the coordinate along
$\bf S^1$ as non-compact. Otherwise we get a quotient of the $\bf
AdS_3$ space.

Now from (\ref{ecomp11}) we see that if we take  $|n|$ large keeping
the other charges fixed, the radius $R$ of the circle $\bf S^1$ becomes
large. Thus we expect that in this limit the entropy will have the
form given in (\ref{ecomp8}) even after inclusion of higher
derivative corrections. However when all charges are of the same
order then the higher derivative corrections to the action will
contain terms which cannot be regarded as the dimensional reduction
of a three dimensional general coordinate invariant action of the
form given in (\ref{ecomp6}), and the higher derivative corrections
to the entropy will cease to be of the form given in (\ref{ecomp8}).
This can be seen explicitly by taking into account the effect of the
four derivative Gauss-Bonnet term in the four dimensional effective
action describing heterotic string compactification on $\bf T^4\times
S^1\times \wt S^1$. The lagrangian density has a term of the form:
\be\label{ecomp13} \Delta \LL= \sqrt{-\det g}\,  \phi(a,S) \,
(R_{\mu\nu\rho\sigma} R^{\mu\nu\rho\sigma} -4 R_{\mu\nu} R^{\mu\nu}
+ R^2)\, , \ee where \be \label{ecomp14} \phi(a,S) = -{3\over
16\pi^2} \ln \left(2 S |\eta(a+iS)|^4\right)\, . \ee Here
$\eta(\tau)$ is the Dedekind eta function and $a$ denotes the axion
field whose near horizon value vanishes for the black hole we are
considering. The effect of (\ref{ecomp14}) on the
black hole entropy can be computed using the entropy function
method, and to first order its effect is to
 give an additive contribution to the entropy of the
form $-2\pi\Delta \LL$ evaluated in the background (\ref{ecomp11}).
This gives \be\label{ecomp15} \Delta S_{BH} = 64 \, \pi^2 \,
\phi(0,S)|_{S=\sqrt{\left| {nw/ \wt N \wt W}\right|}} = -12 \,
\ln\left[ 2\sqrt{\left| {nw\over \wt N \wt W}\right|} \eta\left( i
\sqrt{\left| {nw\over \wt N \wt W}\right|}\right)^4\right]\, . \ee
{}In the limit of large $|n|$ at fixed values of the other charges,
$S$ is large and $\eta(iS)\sim e^{-\pi S /12}$. Thus the leading
correction to $\Delta S_{BH}$ given in (\ref{ecomp15}) goes as
\be\label{e16} 4\pi \sqrt{\left| {nw\over \wt N \wt W}\right|}\, .
\ee Since this is proportional to $\sqrt{|n|}$ we see that the
expression for the entropy retains the form given in (\ref{ecomp8})
with some correction terms in $c_L$, $c_R$.\footnote{For $n<0$,
\i.e.\ non-supersymmetric extremal black holes, the entropy gets
some additional corrections from other higher derivative terms which
further corrects the expression for $c_R$.} However when all the
charges are of the same order then $S$ is of order unity and we
cannot express the corrected entropy $\EE+\Delta\EE$ in the form
given in (\ref{ecomp8}).

It is instructive to study the origin of the terms which break the
SO(2,2) symmetry of $\bf AdS_3$. First of all (\ref{ecomp15}) contains a
correction term proportional to $\ln S\sim \ln\left|
{nw\over \wt N\wt W}\right|$.
This can be traced to the effect of replacing the continuous
integral over the momentum along $\bf S^1$ by a discrete sum. There are
also additional corrections involving powers of $e^{-2\pi S}$. These
can be traced to the effect of Euclidean 5-branes wrapped on
$\bf K3\times S^1\times \wt S^1$\cite{Harvey:1996ir}. Since the 5-brane
has one of its legs along $\bf S^1$, it breaks the SO(3,1) isometry of
Euclidean $\bf AdS_3$.

The above example also illustrates the basic difference between the
approximation scheme used by the $\bf AdS_3$ and $\bf AdS_2$ based
approaches. The $\bf AdS_3$ based approach is useful when we take
the momentum along the $\bf AdS_3$ circle $S^1$ to be large keeping
the other charges fixed. In this limit the size of $S^1$ becomes
large (see eq.(\ref{ecomp11})) and hence the $SO(2,2)$ symmetry of
$\bf AdS_3$ is broken weakly. As a result the entropy has the form
(\ref{ecomp8}). In the CFT living on the boundary of $\bf AdS_3$,
this corresponds to a state with large $L_0$ (or $\bar L_0$)
eigenvalue, keeping the central charge fixed. This is precisely the
limit in which the Cardy formula for the degeneracy of states is
valid. On the other hand the $\bf AdS_2$ based approach is useful if
all the charges are large since in this limit the $\bf AdS_2$ has
small curvature, and we can use the derivative expansion of the
effective action to find a systematic expansion of the entropy and
the entropy function in inverse powers of charges.

It is natural to wonder about possible additional contribution to
the entropy function from other Euclidean brane configurations, {\it
e.g.} heterotic world-sheet instantons. Since in the supergravity
approximation the moduli associated with the $\bf T^4$ part are not
fixed by the attractor mechanism, they can be chosen to have any
value that we like. If we take one of the circles of the torus to be
of sufficiently small size, then the fundamental heterotic string
world-sheet, wrapped on the two dimensional torus spanned by this
circle and the circle $\bf S^1$ that becomes part of $\bf AdS_3$,
can be made to have arbitrarily small action and could in principle
give a large contribution to the effective action. This in turn
would break the $SO(3,1)$ symmetry of Euclidean $\bf AdS_3$
strongly. In this case however it is known that these instantons do
not lift the flat directions associated with the moduli of $\bf
T^4$. Since the near horizon field configuration is obtained by
extremizing the entropy function it follows that the entropy
function cannot receive any contribution from these world-sheet
instantons. As a result the entropy also does not receive any
contribution from such corrections. The key point in this argument
is that the function whose extremization gives the near horizon
geometry also gives the entropy.

We expect this to be a generic situation, namely that even in cases
where the near horizon geometry has an $\bf AdS_3$ factor, for some
choices of the undetermined moduli there are potential sources
for strong breaking of the $SO(3,1)$ symmetry. One then requires
use of non-renormalization theorems which prevent lifting of
flat directions associated with these undetermined moduli, together
with the fact that the extremization of the entropy function
determines all these moduli, to argue that the entropy function does
not receive any correction from these $SO(3,1)$ breaking terms.

In the supersymmetric case the correction to the entropy given in
(\ref{ecomp15}) can be shown to agree with the corresponding result
for statistical entropy\cite{LopesCardoso:2004xf,David:2006yn}.
It will be important to
examine if similar agreement also holds for the non-supersymmetric
extremal black holes.

\subsection{Black holes without $\bf AdS_3$ factor \label{scomp3}}

In the examples discussed above the effect of deviation from the
$\bf AdS_3$ geometry shows up in the non-leading order. However the
non-leading corrections to the entropy, being the analog of finite
size effects, are dependent on the ensemble used to compute the
entropy and could introduce an ambiguity in the definition of the
entropy. A related phenomenon is the breakdown of the thermal
description close to the extremal limit discussed in the paragraph
below (\ref{enta}). The lower limit on the non-extremality
parameter introduced there would give rise to an additional
contribution to the entropy depending on the precise value of the
non-extremality parameter. Such corrections could mask the higher
derivative corrections.\footnote{In principle these ambiguities are
present for both BPS and non-BPS black holes. Nevertheless for the
BPS states the comparison between the statistical entropy and black
hole entropy has been carried out for corrections which are
suppressed by powers of $Q^{-2}$ using a microcanonical
ensemble\cite{LopesCardoso:2004xf,David:2006yn}.
If we really needed to introduce a
non-extremality parameter of order $l_{pl}^2/Q^2$ in order to be
able to calculate the entropy then this would have introduced
additional
corrections to the entropy which are suppressed
by power of $Q^{-2}$
and which depend on the precise value of the non-extremality
parameter. In this case precision comparison between the two
entropies would not have been possible.} Furthermore, for
non-supersymmetric black holes with runaway scalars, there is an
independent need to consider slightly non-extremal black holes (more
discussion on this can be found in \S\ref{discussion} and
\S\ref{srunaway}). For these reasons it will be useful to find
examples where the leading solution itself does not have an $\bf AdS_3$
factor in its near horizon geometry. This will be the subject of
study in this section. In these examples the `leading solution' does
not necessarily mean the solution in the supergravity approximation.
For example some of these examples will involve small black holes
whose leading entropy comes from higher derivative terms.

The first example involves M-theory compactified to five dimensions
on a Calabi-Yau three-fold. Extremal non-rotating black holes in the
five dimensional theory that could be BPS or non-BPS would have near
horizon geometry $\bf AdS_2\times S^3$ and correspond to some
microscopic configuration of M2-branes wrapping the 2-cycles of the
Calabi-Yau space. In the special situation when the Calabi-Yau space
is elliptically fibered with a base $B$, the theory has a dual
description as type IIB compactification on $\bf B \times S^1$, and
the $\bf S^1$ factor can combine with the $\bf AdS_2$ to produce a
locally $\bf AdS_3$ space\cite{Vafa:1997gr}. However, in general,
one can choose a Calabi-Yau manifold that is not elliptically
fibered. In this case there is no duality frame in which the compact
space has a circle factor, and the near horizon geometry of extremal
black holes in the resulting theory does not have an obvious $\bf
AdS_3$ factor.\footnote{It is in principle possible that in some
appropriate limit a contractible circle inside the Calabi-Yau space
becomes large and combines with the $\bf AdS_2$ factor to form an
approximately $\bf AdS_3$ space.} However one can still use entropy
function method to calculate the entropy of these extremal black
holes. It is not known at present how to compute the microscopic
entropy but our conjecture implies a new prediction that it should
equal the macroscopic black hole entropy for both BPS and non-BPS
extremal black holes.

Another example of an extremal black hole without $\bf AdS_3$ factor
is an extremal, non-BPS, electrically charged black hole in
heterotic string theory in ten dimensions. An elementary string with
only right-moving oscillator excitations of level $N_R$ and
left-moving charge vector $\vec Q$, satisfying the level matching
condition $\vec Q^2 = 2 N_R + 1$ (in the Neveu-Schwarz sector),
describes a state that breaks all supersymmetries. The statistical
entropy computed from counting of the degeneracy of states is given
by \be\label{exstat} S_{stat}\simeq 2\sqrt 2 \pi \sqrt{N_R} \simeq
2\pi\sqrt{Q^2} \ee
 for
large $Q^2$. We expect the supergravity description of this state to
be an extremal small black hole. Since there is no physical circle
associated with the charge $\vec Q$, there is no underlying $\bf AdS_3$
geometry. Nevertheless our argument will imply that the microscopic
entropy of the system should match the macroscopic entropy
associated with the small black hole. In fact from the general
scaling argument of
\cite{Sen:1995in,Peet:1995pe,Sen:2005kj} it follows that
the entropy of such a black hole is proportional to $\sqrt{Q^2}$
in agreement with (\ref{exstat}). It
will be interesting to explore if the constant of proportionality
agrees with the prediction from the microscopic entropy.

We could try to find variants of this example in lower dimensions by
considering heterotic string theory on tori. However in this case
there is  a T-duality transformation that maps the original charge
vector $\vec Q$ to momentum and winding along a compact circle. The
small black hole describing this state could have
an underlying $\bf AdS_3$
factor that combines the $\bf AdS_2$ component of the near horizon
geometry, and the circle along which the string carries momentum.
In this case the equality of the macroscopic
and microscopic entropy would
follow from the non-renormalization of the central charge of the
boundary CFT.\footnote{The central charges $c_L$ and $c_R$ of
the boundary CFT, related to the appropriate gauge
and gravitational Chern-Simons terms of the bulk theory,
has been carried out only for
five-dimensional black strings \i.e.\ four dimensional
black holes. It would be interesting to
do this computation for higher dimensional small black holes
and verify that the central charges $c_L$ and $c_R$ of the
boundary theory agree with the central charges of the fundamental
heterotic string world-sheet theory.}

We can however find extremal small black holes
without $\bf AdS_3$ near horizon geometry by considering heterotic
string theory compactified on manifolds without an $\bf S^1$ factor,
{\it e.g.} {\bf K3} or Calabi-Yau three fold. Let us for definiteness
consider heterotic string theory compactified on a Calabi-Yau 3-fold
and consider an elementary heterotic string carrying left-moving
charges and right-moving oscillator excitations satisfying level
matching condition. The statistical entropy of the system is again
given by $2\pi\sqrt{\vec Q^2}$. Again we expect the system to be
described by a small black hole with $\bf AdS_2$ near horizon geometry
but no underlying $\bf AdS_3$. Our arguments will imply that the
macroscopic entropy of this black hole will match the statistical
entropy.

In this case in fact we can give an argument showing that for large
$Q^2$ the macroscopic entropy is also given by $2\pi\sqrt{\vec
Q^2}$, thereby verifying our conjecture. The result is based on a
universality argument similar to the one given in \cite{Sen:1997is}
for BPS black holes. We begin with the observation that since the
string coupling square near the horizon is  of order $1/\sqrt{\vec
Q^2}$, we can carry out our analysis using tree level effective
action. On the other hand the part of the tree level effective
action that is relevant for our computation is the one that involves
the metric, the Maxwell field and the dilaton and is independent of
the manifold on which the theory is compactified. Thus we can
replace the Calabi-Yau three fold by $T^6$ without changing the
result for the microscopic entropy. In this case however the black
hole under consideration can be rotated by T-duality to the one that
carries only momentum and winding along a circle. For this system
there is an $\bf AdS_3$ factor in the near horizon geometry and we
can calculate the entropy using the Kraus-Larsen argument to be
$2\pi\sqrt{ \vec Q^2}$. Thus the small black hole in heterotic
string theory on $T^6$ must also have entropy $2\pi\sqrt{Q^2}$ in
agreement with the microscopic entropy.

The final example we will consider is that of the entropy of
possible small black holes describing fundamental type II strings.
Let us consider an appropriate compactification of type IIA or IIB
string theory down to four non-compact dimensions where the
compactification breaks all the space-time supersymmetries in the
left-moving sector of the world-sheet and preserves at least $\NN=2$
supersymmetry in the right-moving sector. We will also assume that
the compact space contains an $\bf S^1$ factor. Examples of such
compactifications can be found in \cite{Sen:1995ff}. We now consider
an elementary type II string in this theory, wound $w$ times along
$\bf S^1$ and carrying momentum $n$ along $\bf S^1$. For $nw>0$ we
can get extremal BPS states by keeping all the right-moving
oscillators in their ground state and exciting the left-moving
oscillators to level $nw$. On the other hand for $nw<0$ we can get
extremal non-BPS states by keeping all the left-moving oscillators
in their ground state and exciting the right-moving oscillators to
level $|nw|$. For large $|nw|$ the statistical entropy, computed
from the degeneracy of states, is given in both cases by
\be\label{typeii1} S_{stat} = 2\sqrt 2\, \pi \, \sqrt{|nw|}\, . \ee
We would naively expect that in analogy with the heterotic example,
the gravitational description of this system will be a small black
hole. In fact a scaling argument along the line of
\cite{Sen:1995in,Peet:1995pe,Sen:1997is} shows that the string
coupling square at the horizon goes as $1/\sqrt{|nw|}$ so that to
leading order we can consider only the tree level effective action,
and  the contribution to the entropy at tree level, if non-zero,
must be proportional to $\sqrt{|nw|}$. Furthermore, the part of the
tree level effective action relevant for computing the entropy is
invariant under the world-sheet parity transformation to all orders
in the $\alpha'$ expansion since it does not know about the
left-right asymmetry introduced by the compactification. As a result
the constant of proportionality in the expression for the entropy
must be the same for both BPS and the non-BPS black holes. Thus if
the agreement between the statistical and macroscopic entropy holds
for  extremal BPS black holes, it must also hold for extremal
non-BPS black holes.

If as in heterotic string theory we proceed with the assumption that
the $\bf AdS_2$ factor of the near horizon geometry combines with
the $\bf S^1$ factor to give a locally $\bf AdS_3$ space, we run
into inconsistent results. Essentially the coefficients of the
relevant Chern-Simons terms vanish in the tree level type II
effective action, and as  result $c_L$ and $c_R$ appearing in
(\ref{ecomp8}) would vanish.Thus we will get vanishing answer for
the entropy in disagreement with the statistical entropy both for
the BPS and the non-BPS systems. Put another way, in this case the
entropy function has no non-trivial extremum where the condition
(\ref{ecomp4}) is satisfied.

The only possible way out seems to be that the entropy function now
has a different extremum at which the condition
(\ref{ecomp4}) is not satisfied. As a result the near horizon geometry
does not have a locally $\bf AdS_3$ factor. If there is indeed such a
non-trivial extremum, then by the general scaling argument the
macroscopic entropy, represented by the value of the entropy
function at this extremum, will be proportional to $\sqrt{|nw|}$.
At present we do not know if the entropy function has such an
extremum, and even if has such an extremum, what would be
the precise coefficient appearing in front of $\sqrt{|nw|}$. All
we can say is that if this procedure leads to a macroscopic entropy
that agrees with the statistical entropy for BPS black holes, then
similar agreement would also be present for extremal non-BPS
black holes.

\subsection{Black holes from black strings \label{scomp2}}

Typically in cases where the near horizon geometry is described by a
locally $\bf AdS_3$ space, the microscopic description of the black hole
involves a string-like object wrapped along an internal circle
$\bf S^1$, where the string itself may be the result of wrapping some
brane configuration on an internal manifold. The charge $n$
conjugate to the electric flux through $\bf AdS_2$ has the
interpretation of momentum carried by the string along the internal
circle.
If in the infrared the world-sheet theory of the string flows to a
conformal field theory with central charges $(C_L,C_R)$ then
for large $|n|$ the statistical entropy of extremal states in this CFT,
carrying only left-moving or only right-moving excitations,
is given by
the Cardy formula: \bea
\label{ecardy} S_{stat} &&= 2\pi \sqrt{ C_L \, n\over 6} \quad
\hbox{for} \, n>0\, ,
\nonumber \\
&&= 2\pi \sqrt{C_R \, |n|\over 6} \quad \hbox{for} \, n<0\, . \eea
Note that in the $(0,4)$ SCFT the states carrying left-moving
momentum ($n>0$) are BPS but states with right-moving momentum
($n<0$) are non-BPS.

Let us now consider two possibilities. Let $\lambda$ denote the
parameter that controls the strength of the interaction in the
world-sheet theory of the string. If $\lambda$ is a marginal
deformation of the CFT
then we can vary it continuously.
(This should correspond
to the case where in the black hole description the attractor
equations leave $\lambda$ undetermined.)
Since in a
two dimensional CFT the central charges do not change under a
marginal deformation, we can compute them for small $\lambda$ by
ignoring all interactions. This will then also give their values at
large $\lambda$ where the black hole description is good.
The second possibility is that $\lambda$ is not a marginal deformation
and that in the infrared it gets fixed to a strong coupling value
so that the dual black hole description has
a horizon geometry with small curvature. In this case however
we cannot calculate the central charges in the microscopic theory
directly.
But for the
special situation when the two dimensional boundary CFT has $(0, 4)$
super conformal symmetry this is possible. The key point is that the
$(0,4)$ world-sheet supersymmetry acting on the right-moving modes
has $SU(2)_R$ R-symmetry. Furthermore supersymmetry relates the
anomaly in the $SU(2)_R$ R-symmetry to the central charge $C_R$.
Thus the calculation of $C_R$ at strong coupling can be related to
the calculation of the $SU(2)_R$ anomaly at strong coupling. The
latter on the other hand is not renormalized beyond one loop. Thus
knowing the perturbative answer for $C_R$ we can calculate $C_R$ and
hence the statistical entropy at strong coupling for
non-supersymmetric extremal black holes. Moreover, the
quantity $C_L-C_R$ is related to the gravitational anomaly of the
world-sheet theory of the string. Hence this is also not
renormalized as we go from weak to strong coupling by the 't Hooft
anomaly matching requirement. Thus we can also calculate
$C_L$, and hence the statistical entropy of supersymmetric
extremal black holes at strong coupling.

The non-renormalization of $C_L$ and $C_R$ as we go from weak to
strong coupling regime shows that the statistical entropy of these
systems do not change as we go from the weak to the strong coupling
regime. As a result we should be able to compare the statistical
entropy computed in the weakly coupled regime to the black hole
entropy computed in the strong coupling regime. These arguments
provide an alternate explanation of why the entropy of an extremal
non-BPS black hole, calculated at strong coupling, should agree with
the statistical entropy computed at weak coupling. It also provides
an alternative explanation of why for large $n$ the number of BPS
states do not change as we go from weak coupling to the strong
coupling region. However this argument is less powerful than the one
based on supersymmetry, since this holds only in the limit of large
$|n|$ when the statistical entropy is determined by the central
charge alone.

One cannot fail to notice the similarity between (\ref{ecomp8}) and
(\ref{ecardy}).  As already noted, using anomaly inflow one can
relate the quantities $c_L$ and $c_R$ appearing in (\ref{ecomp8}) to
the left- and right-moving trace anomalies in the CFT living on the
boundary of $\bf AdS_3$. If one further assumes AdS/CFT
correspondence\cite{Maldacena:1997de} then the CFT living on the
boundary  of $\bf AdS_3$ is the same as the CFT describing the dynamics
of the microscopic theory. This allows us to identify $c_L$ and
$c_R$ with $C_L$ and $C_R$ respectively, and makes the equality of
black hole entropy and statistical entropy manifest.

The arguments presented above require that the microscopic
description  be based on the dynamics of a string-like object, which
may not always be the case. This is what happens for the example
described in \S\ref{scomp3} involving black holes in M-theory
compactified on a non-elliptically fibered Calabi-Yau three fold.
Moreover, even when there is an underlying  string, determination of
the central charges alone is not sufficient if one wishes  to go
beyond the leading asymptotics given by the Cardy entropy. This is
the counterpart of the macroscopic result that the $\bf AdS_3$
description is useful in the limit of large $|n|$, but fails when
all the charges are of the same order. The arguments based on
$\bf AdS_2$ near horizon geometry continues to hold in such cases. Thus
for these examples our conjecture  makes nontrivial predictions
about the relation between weak coupling statistical entropy and
`strong' coupling black hole entropy  which would be interesting to
verify.

\section{Rotating black holes} \label{srotate}

Extremal spinning black holes also display attractor behavior which
can be understood from the existence of the underlying entropy
function\cite{Astefanesei:2006dd}. Thus we expect the agreement
between microscopic and macroscopic entropy to hold even in the case
of spinning black holes.

An example of this may be constructed as follows. Let us consider
the D1-D5 system with momentum considered in \S\ref{fivedcount} and
add equal angular momentum in the two planes transverse to the
D5-brane. Since for negative $n$ the system was not supersymmetric
to begin with, it will be non-BPS even after we add angular
momentum. The entropy of this black hole can be computed directly,
but can also be related to the entropy of a four dimensional black
hole\cite{Gaiotto:2005gf,Gaiotto:2005xt} by taking the space
transverse to the brane to be Taub-NUT space. This has the effect of
compactifying an additional dimension (say $x^4$) with the angular
momentum interpreted as the momentum along $x^4$. Since the presence
of the Taub-NUT space does not affect the structure of the black
hole horizon in the limit where the size of the Taub-NUT space is
large, in this limit the black hole entropy will be given by that of
the rotating five dimensional black hole. On the other hand if we
take the Taub-NUT space to be of small size then it is more
appropriate to regard the black hole as a four dimensional black
hole and the entropy will be given by the entropy of a four
dimensional black hole carrying momentum along the $x^4$ direction.
This is precisely the system described in \S\ref{fourdcount}.
Since the entropy cannot depend on the size of the Taub-NUT which,
being the asymptotic radius of $x^4$, is one of the moduli, we see
that the entropies of the five and four dimensional black holes must
be identical. On the other hand the microscopic counting of the four
and the five dimensional systems are almost identical, with the four
dimensional system receiving some additional contribution from the
dynamics of the Taub-NUT space and the motion of the D1-D5 system in
the Taub-NUT background\cite{David:2006yn}. However these
contributions are subleading and do not affect the leading entropy
in the limit of large charges. Thus the black hole entropy of the
five dimensional rotating black hole agrees with the statistical
entropy of the same system as a consequence of the corresponding
agreement for the four dimensional
non-rotating system discussed in
\S\ref{fourdcount}.

A closely related example is as follows \footnote{
A discussion for the attractor mechanism being the basis of the agreement between the
microscopic and macroscopic entropy in this example also appears in the forth-coming paper
\cite{AGM}.}\cite{Emparan:2006it}.
Let us consider type IIA
string theory compactified on $\bf \MM$, where
$\MM$ can be $\bf K3\times T^2$, $\bf T^6$ or a Calabi-Yau 3-fold,
and take a system of $q_0$
D0-branes and one D6 brane in this theory. Using the duality between
type IIA string theory and M-theory on $\bf S^1_M$, this configuration
lifts to M-theory on Taub-NUT space $\times \MM$
with $q_0$ units of momentum flowing along the asymptotic
circle $\bf S^1_M$ of the Taub-NUT space. If the asymptotic radius of
the $M$ -theory circle $S_M^1$  is big, then the center of Taub-NUT
space is approximately flat $4+1$ dimensional space-time $R^{4,1}$.
The D0-brane charge $q_0$
now can be interpreted as equal angular momentum
along a pair of orthogonal planes in $R^4$ and we get a
neutral extremal rotating black hole sitting in the
center of this approximately flat $4+1$ dimensional space-time. If
we denote the rotation group $SO(4)$ of $R^{4,1}$ by $SU(2)_L \times
SU(2)_R$, then the black hole has $q_0$ units of angular momentum
lying in $SU(2)_L$.

This system breaks all supersymmetries. The Bekenstein-Hawking
entropy of the black hole can be easily computed and is given by,
\be \label{BH} S_{BH}= \pi |q_0|. \ee We can in fact consider a more
general class of black holes which also carry angular momentum $J_R$
associated with $SU(2)_R$. {}From the five dimensional viewpoint
this would correspond to having unequal angular momentum along the
two orthogonal planes of $R^4$. From the four dimensional viewpoint
this describes an extremal charged rotating black hole. The entropy
of the corresponding black hole can also be computed easily and
yields the answer
\be\label{BHnew} S_{BH} = 2\pi
\sqrt{\frac{(q_0)^2}{4} - J_R^2}\, .
\ee

For $\MM=\bf T^6$ the microscopic entropy of this system was
computed in \cite{Emparan:2006it} by studying the dynamics of the
D0-D6 system and yields the answer
\be\label{BHmicro} S_{stat} =
2\pi \sqrt{\frac{(q_0)^2}{4} - J_R^2}\, ,
\ee
in agreement with the
Bekenstein-Hawking entropy.\footnote{We expect that a similar
computation can be done at least for $\MM=\bf K3\times T^2$.} Thus
this provides an example where the macroscopic entropy of an
extremal non-BPS black hole calculated at `strong' coupling agrees
with the statistical entropy of the system calculated at weak
coupling.

In this system, the initial configuration, when interpreted as a
rotating black hole solution in five dimensions, does not have an
obvious $\bf AdS_3$ factor. However interpreted as a four
dimensional black hole this system is dual to heterotic string
theory on $T^6$ for $\MM=\bf K3\times T^2$ and type IIA string
theory on $T^6$ for $\MM=\bf T^6$. Let us for definiteness
concentrate on the case $\MM= \bf K3\times T^2$; the case for
$\MM=\bf T^6$ may be analyzed in a similar manner. If we set $J_R=0$
then black hole solution describes a non-rotating black hole in four
dimensions carrying some electric and magnetic charges $(Q,P)$ with
$Q^2 = P^2=0$, $Q\cdot P = q_0^2$. Since in the supergravity
approximation the entropy is a function of the duality invariant
combination $D\equiv [P^2 Q^2 - (P\cdot Q)^2]/4$, we can calculate
the entropy by choosing a different representative with the same
value of $D$ that has an $\bf AdS_3$ factor in its near horizon
geometry. For example one can map this system to the familiar
D1-D5-KK5-$\bar{\rm P}$ system of type I theory discussed in
\S\ref{fourdcount}. The entropy of this system is given by
\be\label{dualinv} S_{BH}=2 \pi \sqrt{-D} = \pi |q_0|\ee in
agreement with (\ref{BH}). This allows us to use an $\bf AdS_3$
based argument along the line of \S\ref{scomp1} for explaining the
agreement between the statistical and black hole entropy.

Note however that if we begin with a rotating extremal black hole
configuration in M-theory on $\bf K3\times
T^2 \times  {\rm Taub-NUT}$
where $\bf K3$, $\bf T^2$ and Taub-NUT
have sizes large compared to the
11-dimensional Planck scale, and the angular momentum is large so
that the horizon size is large compared to the Planck scale, then
the original description in terms of M-theory is a weakly coupled
description. A duality transformation that takes this to a system
with an $\bf AdS_3$ factor in the near horizon geometry must map it to a
region of the moduli space where some degrees of freedom in the
final description are strongly coupled since we cannot have two
different weakly coupled descriptions of the same background. This
could break the $SO(3,1)$ symmetry of Euclidean $\bf AdS_3$ strongly by
the various mechanisms discussed in \S\ref{scomp1}. In order to get
an $\bf AdS_3$ geometry with weakly coupled degrees of freedom, we must
begin at a corner of the moduli space where the original description
in terms of M-theory is strongly coupled. We then need to
invoke the
attractor mechanism to argue that the entropy of the system does not
change as we move from the weakly coupled region to the  strongly
coupled region. The  implicit use of attractor mechanism can also be
see from that fact that in order to argue that the entropy is  a
function only of $P^2 Q^2 - (P\cdot Q)^2$ without doing explicit
calculation, we need to assume that it does not depend on the
asymptotic moduli. Otherwise we could construct more general duality
invariant combinations of moduli and charges on which the entropy
could depend.

Let us now consider the effect of switching on $J_R$ in the original
D0-D6 system. {}From the point of view of a (3+1) dimensional theory
this corresponds to imparting an angular momentum on the system. The
effect of this is to change (\ref{dualinv}) to
\cite{Rasheed:1995zv,Larsen:1999pp,
Cvetic:1996kv,Astefanesei:2006dd} (see eqs.(5.104), (5.105) of
\cite{Astefanesei:2006dd}) \bea \label{ecomp8n} S_{BH} && = 2\pi
\sqrt{D+J_R^2} \quad \hbox{for} \, D+J_R^2>0\, ,
\nonumber \\
&& = 2\pi\sqrt{-D-J_R^2} \quad \hbox{for} \, D+J_R^2<0\, . \eea The
case $D+J_R^2<0$ corresponds to the branch of the rotating D0-D6
black hole which has no ergo-sphere, while the case $D+J_R^2>0$
corresponds to the branch with an
ergo-sphere\cite{Rasheed:1995zv,Larsen:1999pp,Astefanesei:2006dd}.
Substituting $D=-(q_0)^2$ in the second equation of (\ref{ecomp8n})
we recover eq.(\ref{BHnew}).

Let us now return the case where $\MM$ is a Calabi-Yau 3-fold. First
suppose  $\MM$ is elliptically fibered with base $\bf B$. Then by
the usual M-theory - F-theory duality we can relate this to IIB on
$\bf B\times S^1$. After performing the complicated set of duality
transformations described earlier we can arrive at a configuration
where the near horizon geometry has an $\bf AdS_3$ factor, and the
black hole entropy can be calculated using eqs.(\ref{ecomp8}) as
usual. However if the manifold $\MM$ is not elliptically fibered
then there is no obvious way at least to associate an $\bf AdS_3$
space with the compactification, and therefore we cannot apply
eq.(\ref{ecomp8}) to compute the entropy. However a discussion
analogous to that of \cite{Astefanesei:2006dd} will apply for the
five dimensional rotating black hole, showing that the attractor
mechanism does work in this case as well. And thus the arguments
presented in   this paper will provide a prediction for the
microscopic counting of states.

\section*{Acknowledgments}

We would like thank Rudra Jena, Per Kraus, Shiraz Minwalla, Suresh
Nampuri, and Masaki Shigemori for useful discussions. A.D. would
like to thank the Aspen Center for Physics and LPTHE where part of
this work was completed. S.P.T. acknowledges support from the
Swarnajayanti Fellowship, Grant No. ZTHSJ XP-104, D.S.T, Govt. of
India and support from the D.A.E, Govt. of India. Most of all we
thank the people of India for supporting research in String Theory.

\bibliography{nonsusyref}
\bibliographystyle{jhep}

\end{document}